\documentclass[preprint,review,12pt,3p,times]{elsarticle}

\usepackage{ifxetex}
\usepackage{ifpdf}
\ifxetex
  \usepackage[bookmarksnumbered,colorlinks,pagebackref]{hyperref}
\else
  \usepackage[CJKbookmarks=true,unicode,bookmarksnumbered,colorlinks,pagebackref]{hyperref}
\fi

\usepackage{amssymb}
\usepackage{amsmath}
\usepackage{amsthm}
\usepackage{bm}
\usepackage{color}
\usepackage{epstopdf}
\usepackage{subfig}
\usepackage[ruled]{algorithm2e}
\usepackage{algorithm2e,setspace}
\usepackage{upgreek}
\usepackage{booktabs}
\usepackage{array}
\usepackage{makecell}
\usepackage{lineno}
\usepackage{setspace}
\usepackage{upgreek}

\biboptions{sort&compress}

\newcommand{\refeq}[1]{(\ref{#1})}  

\journal{}

\begin{document}

\begin{frontmatter}



\title{A generalized three-dimensional hybrid contact method for smoothed particle hydrodynamics}

\author[BIT]{Wenbin Liu\corref{cor1}}
\author[BIT]{Zhuoping Duan}
\author[BIT]{Yan Liu}
\author[BIT]{Fenglei Huang}

\address[BIT]{State Key Laboratory of Explosion Science and Technology, Beijing Institute of Technology, Beijing 100081, China}

\cortext[cor1]{Author to whom correspondence should be addressed: nxliuwenbin@163.com}

\begin{abstract}
    When the effects of relative motion at the solid object interfaces are not negligible, the contact method is required in the smoothed particle hydrodynamics (SPH) method to prevent virtual shear and tensile stresses. However, there is still a lack of a three-dimensional (3D) contact method that can be well applied to various deformation situations, especially for extreme deformation. In this study, we propose a generalized 3D hybrid contact method for SPH. First, an improved high accuracy free-surface particle detection method is developed, including optimization of the detection process to reduce the detection time and consideration of the effect of material compressibility on the filtering parameters to extend the existing semi-geometric method from the incompressible (weakly-compressible) field to the compressible field. Then, a novel 3D local surface reconstruction method is developed based on the free-surface particles and region growing method, including the selection of the initial edge, the principle of triangle expansion, and the evaluation function, followed by the surface-surface contact detection and enforcement of normal penalty and tangential friction forces according to the penalty function method and Coulomb friction law. Finally, the particle-particle contact method is added to deal with cases where surface-surface contact fails, e.g., some particles are unable to reconstruct the local surface when the material undergoes extreme deformations. The proposed method is validated by several numerical tests, and the results show that the proposed method is capable of handling various contact problems with accuracy and stability, including small, large, and extreme deformation problems.
\end{abstract}




\end{frontmatter}

\section{Introduction}\label{sec:intro}

Smoothed particle hydrodynamics (SPH) is a representative particle method \cite{Lucy1977, GingoldMonaghan1977} first proposed for solving astrophysical problems. Due to its Lagrangian and meshfree properties, the SPH method has been applied to a variety of scientific and engineering problems \cite{Monaghan2012, ZhangSun2017, YePan2019, LiberskyPetschek1993, GrayMonaghan2001, ColagrossiLandrini2003, SaitohMakino2013, SunTouze2021, LiuMa2021, ZhangLong2019, VyasCummins2021, RahimiMoutsanidis2022}, such as astrophysics, hydrodynamics, and explosion and impact dynamics.

As the application fields continue to expand, the SPH method is limited by some challenging problems, and the contact boundary condition is one of them. Unlike mesh-based methods, the lack of connectivity of the particles makes it difficult to accurately determine the boundary, especially in the case of extreme deformations, making the treatment of the contact boundary quite difficult. In the early applications of the SPH method, the contact boundary condition was treated by the momentum conservation equation \cite{LiberskyPetschek1993}, i.e., the particles of one object could be considered as neighboring particles of another object and were taken into account in the particle approximation. When this method is utilized to address contact interactions between solid objects, the consideration of the stress tensor effects of other objects in the momentum equation introduces virtual tensile and shear stresses, which produce unacceptable results when the effects of relative motion at the object interfaces are not negligible.

Two classes of methods have been proposed to overcome the contact boundary condition: the particle-particle method and the surface-surface method. In the particle-particle method \cite{CampbellVignjevic2000, SeoMin2006, SeoMin2008}, particles are regarded as circles and spheres with certain sizes in two and three dimensions, respectively. It detects the contact by calculating the penetration between circles/spheres and prevents particles from penetrating each other with the penalty function method. This method is simple and efficient to implement. However, the implementation of this method requires the evaluation of the normal vector and is therefore dependent on the distribution of the particles, in addition to the fact that under large deformations, the material surface is prone to the formation of gaps whereby unphysical infiltration occurs. In the surface-surface method \cite{WangChan2014, WangWu2013, XiaoHu2013, ZhangPeng2020, MuTang2023, ZhouFang2022, XiaoLiu2023, XiaoLiu20232, WiragunarsaZuhal2024}, whether contact occurs is determined by establishing particle-segment contact pairs, and if contact occurs, normal and frictional contact forces are imposed, respectively. Wang et al. \cite{WangChan2014} developed a two-dimensional (2D) particle-segment contact method and successfully applied the method to simulate the local deformation behavior of materials under soil-structure interaction. Xiao and Liu \cite{XiaoLiu2023} presented a 2D surface-surface contact method. In this method, object surfaces are reconstructed using the $\alpha$-shape method. The contact between the objects is detected, and the contact boundary conditions are applied. Mu et al. \cite{MuTang2023} developed a 2D hybrid contact method for modeling frictional slips between pre-existing fracture surfaces. The method utilizes the particle-segment contact and particle-particle contact for contact of particles on the contact surfaces and particles at the corners, respectively.  Currently, most of the surface-surface contact methods are only applicable to 2D aspects, while there are fewer three-dimensional (3D) methods. Xiao and Liu \cite{XiaoLiu20232} proposed a 3D surface-surface contact method. In this method, the 3D object surface represented by SPH particles is reconstructed by the global Delaunay triangulation. Based on the reconstructed surface, the contact between different objects is detected, and the contact boundary conditions are applied. Computational results show that the method is applicable to various contact problems and good results are obtained. However, the 3D surface reconstruction method comes from commercial software with an unknown source code, which is not conducive to the generalization and popularization of the method. In addition, when the material undergoes extreme deformation, some free-surface particles are unable to reconstruct the surface, at which time the surface-surface contact fails, which is also an unsolved problem.

In this study, we propose a generalized 3D hybrid contact method for SPH. First, an improved high accuracy free-surface particle detection method is developed, including optimization of the detection process to reduce the detection time and consideration of the effect of material compressibility on the filtering parameters to extend the original semi-geometric method from the incompressible (weakly-compressible) field to the compressible field. Then, a novel 3D local surface reconstruction method is developed based on the free-surface particles and region growing method, including the selection of the initial edge, the principle of triangle expansion, and the evaluation function, followed by the surface-surface contact detection and enforcement of normal penalty and tangential friction forces according to the penalty function method and Coulomb friction law. Finally, the particle-particle contact method is added to deal with cases where surface-surface contact fails, e.g., some particles are unable to reconstruct the local surface when the material undergoes large deformations. The accuracy and stability of the proposed method are first examined by simulating the sliding of the block along the slope and subsequently employed to simulate the ignition response of high explosives and high velocity impact.

This paper is organized as follows. The SPH method is first presented in Section~\ref{sec:2}. Then, the details of the 3D hybrid contact method are provided in Section~\ref{sec:3}. Several numerical applications are given in Section~\ref{sec:4}. The conclusions are presented in Section~\ref{sec:5}.

\section{SPH method}\label{sec:2}

\subsection{Governing equations}

The governing equations considering the material strength are as follows:
\begin{equation}\label{eq:governing-equations}
  \left\{
    \begin{array}{l}
      \displaystyle\frac{\mathrm{d}\rho}{\mathrm{d}t}=-\rho \frac{\partial U^{\beta}}{\partial x^{\beta}}\medskip\\
      \displaystyle\frac{\mathrm{d}U^{\alpha}}{\mathrm{d}t}=\frac{1}{\rho}\frac{\partial \sigma ^{\alpha \beta}}{\partial x^{\beta}}\medskip\\
      \displaystyle\frac{\mathrm{d}E}{\mathrm{d}t}=\frac{\sigma ^{\alpha \beta}}{\rho}\frac{\partial U^{\alpha}}{\partial x^{\beta}}
  \end{array},
  \right.
\end{equation}
where $ \rho $ denotes the density, $ E $ denotes the specific internal energy, $ U^{\alpha} $ denotes the velocity component, $ x^{\beta} $ denotes the spatial coordinate component, $ \sigma ^{\alpha \beta} $ denotes the stress tensor component.

\subsection{SPH equations}

After discretizing Eq.~\refeq{eq:governing-equations} using the kernel and particle approximations, the SPH equations is expressed as
\begin{equation}\label{eq:sph-equations}
  \left\{
  \begin{array}{l}
    \displaystyle \frac{{\rm {d}} \rho_i}{{\rm {d}} t}
    =
    \rho_i \sum\limits_{j = 1}^N {\frac{m_j}{\rho_j} \left( {U}_{i}^{\beta} - {U}_{j}^{\beta} \right)} \frac{\partial W_{ij}^{C}}{ \partial {x}_{i}^{\beta}} \medskip \\
    \displaystyle \frac{{\rm {d}} {U}_{i}^{\alpha}}{{\rm {d}} t}
    =
    \sum\limits_{j = 1}^N{m_j \left( \frac{{\sigma}_{i}^{\alpha \beta}}{\rho _{i}^{2}} + \frac{{\sigma}_{j}^{\alpha \beta}}{\rho_{j}^{2}} - \Pi_{ij} \right) \frac{\partial W_{ij}^{C}}{\partial {x}_{i}^{\beta}}} \medskip \\
    \displaystyle \frac{\mathrm{d}E_i}{\mathrm{d}t}
    =
    \frac{1}{2}\sum_{j=1}^N{m_j\left( \frac{\sigma _{i}^{\alpha \beta}}{\rho _{i}^{2}}+\frac{\sigma _{j}^{\alpha \beta}}{\rho _{j}^{2}}-\Pi _{ij} \right) \left( U_{j}^{\beta}-U_{i}^{\beta} \right) \frac{\partial W_{ij}^{C}}{\partial x_{i}^{\beta}}}
  \end{array},
  \right.
\end{equation}
where $ W $ denotes the kernel function, and the Wendland kernel function \cite{Wendland1995} is used in this study.

The artificial viscosity \cite{Monaghan1989} is used as follows
\begin{equation}
  \Pi_{ij}
  =
  \left\{
  \begin{array}{ll}
    \displaystyle \frac{-\alpha_{\Pi} \bar{c}_{ij} \phi_{ij} + \beta_{\Pi} \phi _{ij}^{2}}{\bar{\rho}_{ij}},
    &
    \left( \boldsymbol{U}_i - \boldsymbol{U}_j \right) \cdot \left( \boldsymbol{x}_i - \boldsymbol{x}_j \right) < 0 \medskip \\
    \displaystyle 0,
    &
    \left( \boldsymbol{U}_i - \boldsymbol{U}_j \right) \cdot \left( \boldsymbol{x}_i - \boldsymbol{x}_j \right) \geqslant 0
  \end{array},
  \right.
\end{equation}
where
\begin{equation}
  \phi_{ij}
  =
  \displaystyle \frac{\bar{h}_{ij}\left( \boldsymbol{U}_i-\boldsymbol{U}_j \right) \cdot \left( \boldsymbol{x}_i-\boldsymbol{x}_j \right)}{\left( \boldsymbol{x}_i-\boldsymbol{x}_j \right) ^2+0.01 \bar{h}^2_{ij}},
\end{equation}
\begin{spacing}{0.8}
\end{spacing}
\begin{equation}
  \bar{c}_{ij}
  =
  \frac{1}{2}\left( c_{i}+c_{j} \right) ,
\end{equation}
\begin{spacing}{0.5}
\end{spacing}
\begin{equation}
  \bar{\rho}_{ij}
  =
  \displaystyle \frac{1}{2}\left( {\rho}_i + {\rho}_j \right),
\end{equation}
\begin{spacing}{0.5}
\end{spacing}
\begin{equation}
  \bar{h}_{ij}
  =
  \displaystyle \frac{1}{2}\left( h_i + h_j \right),
\end{equation}
In above equations, $ c $ denotes the sound speed, and $ h $ denotes the smoothing length.

In order to improve the computational accuracy, the kernel gradient correction technique \cite{ShaoLi2012} is employed as follows:
\begin{equation}
  \left\{
  \begin{array}{l}
	\nabla_i W_{ij}^{C}
    =
    \mathbf{L}_i \nabla_i W_{ij} \medskip \\
	\mathbf{L}_i
    =
    \displaystyle \left[ \sum\limits_{j = 1}^N{\left( \boldsymbol{x}_j - \boldsymbol{x}_i \right) \otimes \nabla_i W_{ij} V_j} \right] ^{-1}
  \end{array}.
  \right.
\end{equation}

The kick-drift-kick method \cite{Monaghan2005} is employed to solve Eq.~\refeq{eq:sph-equations}. First, the velocity at the half time step is:
\begin{equation}\label{eq:velocity-half-step}
  \boldsymbol{U}_{i}^{n + \frac{1}{2}}
  =
  \boldsymbol{U}_{i}^{n}
  +
  \frac{1}{2} \Delta t \left( \displaystyle \frac{ \mathrm{d} \boldsymbol{U}_i}{\mathrm{d} t} \right) ^n,
\end{equation}
Subsequently, the density, specific internal energy, and position at the new time step are updated using the updated velocity in Eq.~\refeq{eq:velocity-half-step} as
\begin{equation}
  \rho_{i}^{n+1}
  =
  \rho_{i}^{n}
  +
  \Delta t \left( \displaystyle \frac{ \mathrm{d}\rho _i}{\mathrm{d}t} \right)^{n+\frac{1}{2}},
\end{equation}
\begin{equation}
  E_{i}^{n+1}
  =
  E_{i}^{n}
  +
  \Delta t \left( \displaystyle \frac{ \mathrm{d} E_i}{\mathrm{d} t} \right)^{n + \frac{1}{2}},
\end{equation}
\begin{equation}
  \boldsymbol{x}_{i}^{n + 1}
  =
  \boldsymbol{x}_{i}^{n}
  +
  \Delta t \boldsymbol{U}^{n + \frac{1}{2}},
\end{equation}
Finally, the velocity at the new time step is updated
\begin{equation}
  \boldsymbol{U}_{i}^{n + 1}
  =
  \boldsymbol{U}_{i}^{n + \frac{1}{2}}
  +
  \frac{1}{2} \Delta t \left( \frac{\mathrm{d}\boldsymbol{U}_i}{\mathrm{d}t} \right) ^{n+1}.
\end{equation}

\section{3D hybrid contact method}\label{sec:3}

In this section, we propose a generalized 3D hybrid contact method for SPH. First, an improved high accuracy free-surface particle detection method is developed based on previous work \cite{LiuMa2023}, including optimization of the detection process to reduce the detection time and consideration of the effect of material compressibility on the filtering parameters to extend the original semi-geometric method from the incompressible (weakly-compressible) field to the compressible field. Next, a novel 3D local surface reconstruction method is developed based on the free-surface particles and region growing method, including the selection of the initial edge, the principle of triangle expansion, and the evaluation function. Then, the 3D surface-surface contact detection is performed based on the reconstructed local surface, and if contact occurs, normal penalty force and tangential friction force are applied according to the penalty function method and Coulomb friction law, respectively. Finally, supplementing the 3D particle-particle contact method to the 3D hybrid contact method for cases where surface-surface contact fails, e.g., some particles are unable to reconstruct the local surface when the object undergoes large deformations. The main operations of the proposed 3D hybrid contact method are presented in Algorithm~\ref{algo:3DHybConMethod}.

\begin{algorithm}
\setstretch{1.3}
\SetAlgoLined
\DontPrintSemicolon

Update particle velocity and position without considering contact forces

\For{$ k $ {\rm over all contact}}
{
  \For{$ m $ {\rm over all particles of master object}}
  {
    \If{{\rm there are particles of other objects in the neighborhood}}
    {
      {\rm Reconstruct local surface}
    }
  }
}

\For{$ k $ {\rm over all contact}}
{
  \For{$ m $ {\rm over all particles of slave object}}
  {
    Find the nearest particle of master object

    Detect of contact

    \If{{\rm contact occurs}}
    {
      Detect of surface-surface contact

      \uIf{{\rm surface-surface contact occurs}}
      {
        Enforcement of surface-surface contact
      }
      \Else
      {
        Enforcement of particle-particle contact
      }
    }
  }
}
\caption{3D hybrid contact method}
\label{algo:3DHybConMethod}
\end{algorithm}

\subsection{Improved high accuracy free-surface particle detection method}

\begin{figure}[htbp]
\centering
\includegraphics[width=8cm]{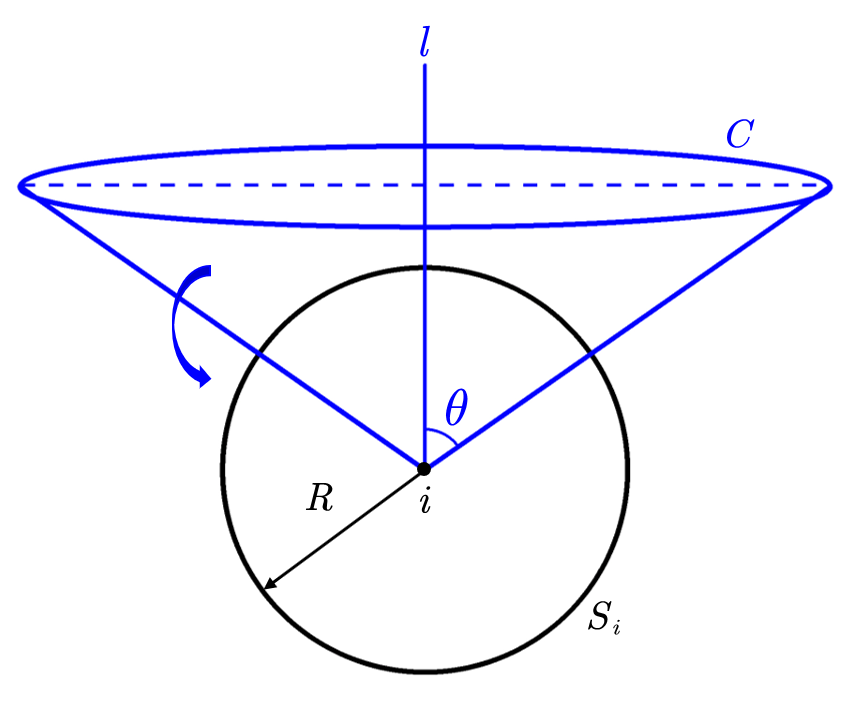}
\captionsetup{font={normalsize,stretch=1.0}}
\caption{Schematic of the geometric method for high accuracy free-surface particle detection. Continuous global scanning is performed through a cone region within the particle neighborhood. If there exists a cone region with no neighboring particles, the particle is identified as a free-surface particle.}
\label{fig:Figure-3-1-1}
\end{figure}

In our previous work \cite{LiuMa2023}, we developed a high accuracy free-surface particle detection method, which performs continuous global scanning through a cone in the neighborhood, and the particle is identified as a free-surface particle if there exists a cone region with no neighboring particles, as shown in Fig.~\ref{fig:Figure-3-1-1}. Tests show that the detection accuracy of the method is improved due to the enhanced ability to detect concave surfaces and the fact that the detection accuracy is independent of the normal vector. However, the method is still time-consuming despite the use of positional divergence for initial filtering, and in addition, the method is only applicable to incompressible (weakly compressible) fields.

An improved high accuracy free-surface particle detection method is developed based on the original method, including optimization of the detection process to reduce the detection time and consideration of the effect of material compressibility on the filtering parameters to extend the original semi-geometric method from the incompressible (weakly-compressible) field to the compressible field. The improved method includes three steps.

\begin{figure}[htbp]
\centering
\includegraphics[width=8cm]{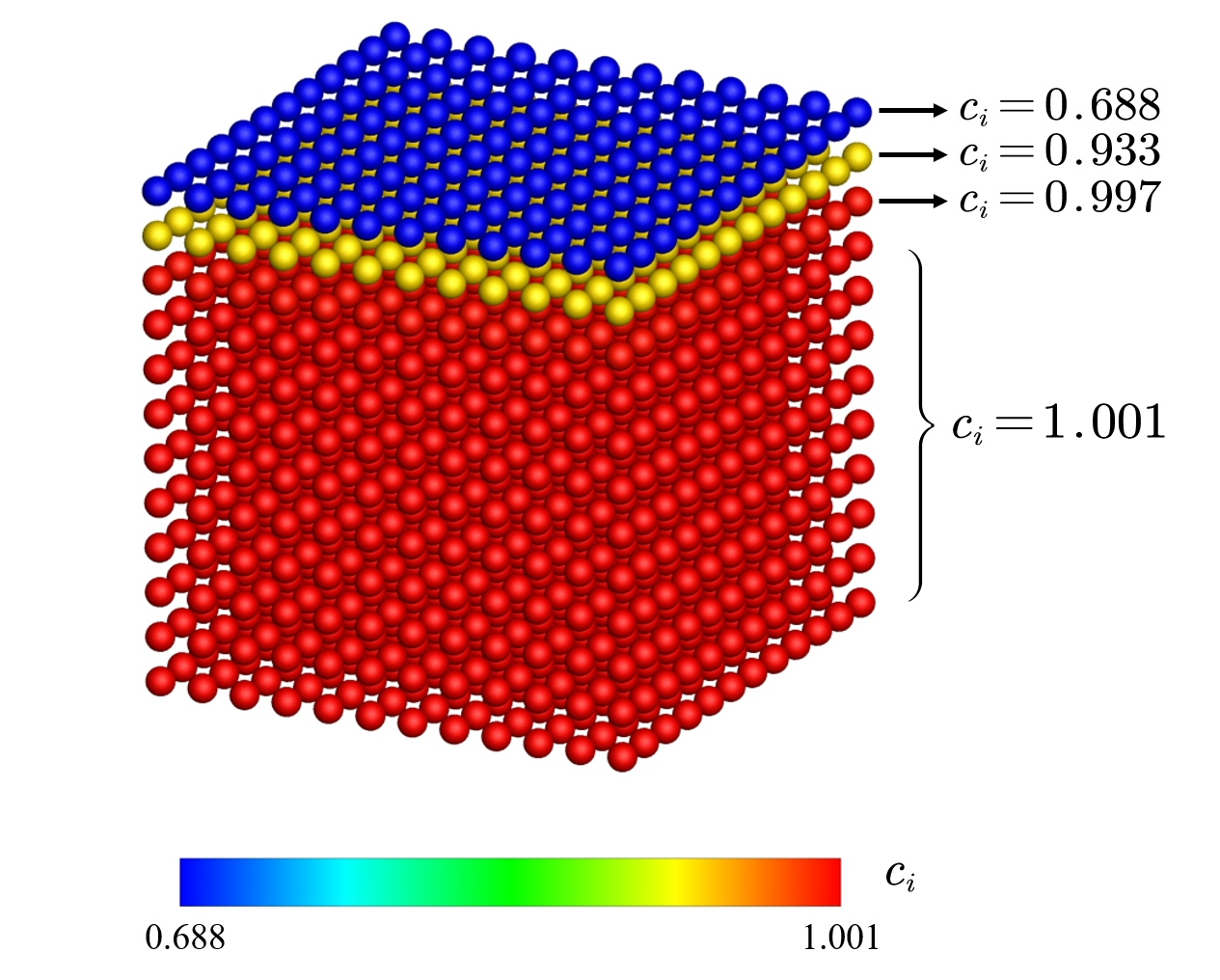}
\captionsetup{font={normalsize,stretch=1.0}}
\caption{The color function of particles in 3D.}
\label{fig:Figure-3-1-2}
\end{figure}

In the first step, the particles adjacent to the free surface are obtained by the color function, which greatly reduces the detection time, and the color function $ c_i $ is calculated as follows:
\begin{equation}
  c_i
  =
  \displaystyle \sum_{j = 1}^N{W_{ij} V_j},
\end{equation}
Figure~\ref{fig:Figure-3-1-2} gives the color function of particles in 3D. In the improved method, we choose 0.96 and 0.60 as the thresholds for direct determination of inner and free-surface particles, respectively. For particles with a color function less than 0.96 but greater than 0.60, a further detection will be performed.

In the second step, a single cone scanning along the normal direction is added to the original method. If there is no neighboring particle in this cone region, the particle is identified as a free-surface particle, otherwise, the global scanning is carried out in the third step. The introduction of the second step also significantly reduces the detection time. The normal direction $ \boldsymbol{n}_i $ is evaluated by the finite particle method \cite{LiuMa2021, LiuLiu2006}.

In the third step, a purely geometric detection is performed on the remaining particles from the second step, specifically, continuous global scanning is performed through a cone region within the particle neighborhood. If there exists a cone region with no neighboring particles, the particle is identified as a free-surface particle. Details of the geometric method are given in previous work \cite{LiuMa2023}.

The final free-surface particle detection expression is as follows.
\begin{equation}
  \left\{
  \begin{array}{lll}
    c_i < 0.60
    &
    \Rightarrow
    &
    i \in FR \medskip \\
	0.60 \leqslant c_i \leqslant 0.96\ \& \&\ C_n = \varnothing
    &
    \Rightarrow
    &
    i\in FR \medskip \\
	0.60 \leqslant c_i \leqslant 0.96 \ \& \&\ C_n \ne \varnothing \ \& \&\ \exists C = \varnothing
    &
    \Rightarrow
    &
    i \in FR \medskip \\
    \mathrm{otherwise}
    &
    \Rightarrow
    &
    i\notin FR
  \end{array}
  \right.
\end{equation}
where $ C_n $ denotes the cone along the normal vector, and $ FR $ denotes the free-surface region.

\begin{figure}[htbp]
\centering
\includegraphics[width=10cm]{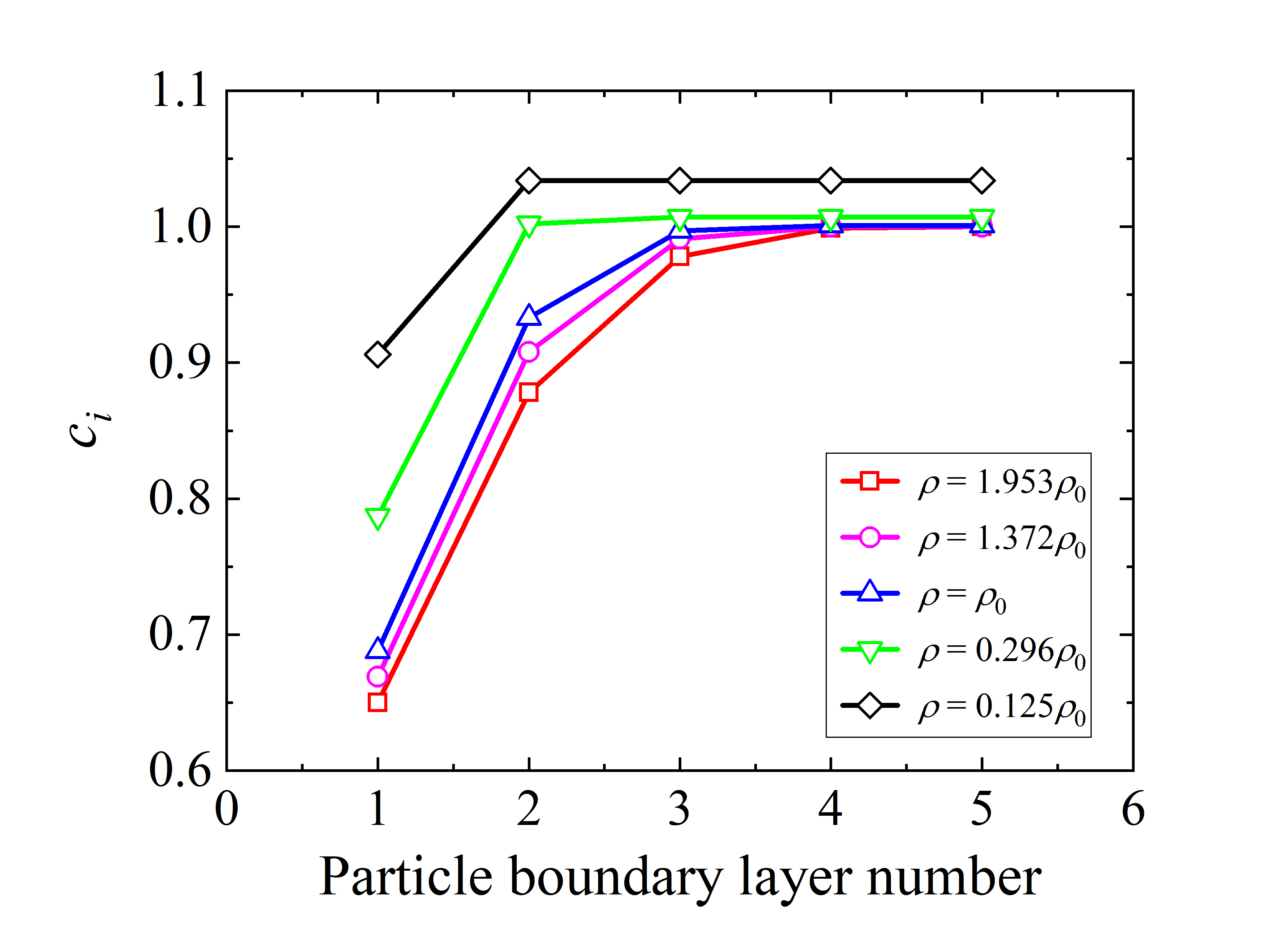}
\captionsetup{font={normalsize,stretch=1.0}}
\caption{The color function of particles adjacent to the free surface in different tension and compression conditions.}
\label{fig:Figure-3-1-3}
\end{figure}

In the field of explosion and impact dynamics \cite{LiberskyPetschek1993, ZhangLiu2017, LiuXi2022, ChenFeng2023}, materials tend to undergo large deformations accompanied by large density variations, which affect the filtration parameters of the color function in the first step. Figure~\ref{fig:Figure-3-1-3} gives the color function of particles adjacent to the free surface in different tension and compression conditions, it can be seen that the distribution of the color functions of particles changes significantly with the material density, which is caused by the drastic change in the number of neighboring particles due to the smoothing length remaining constant. Therefore, the variable smoothing length \cite{Benz1990} is added:
\begin{equation}
  \frac{{\rm {d}} h_i}{{\rm {d}} t}=-\frac{1}{d}\frac{h_i}{\rho _i}\frac{{\rm {d}} \rho _i}{{\rm {d}} t},
\end{equation}
This formula updates the smoothing length with the change of the material density to avoid large changes in the color function of particles, and it also benefits in improving the accuracy of SPH equations.

\subsection{Local surface reconstruction method}

Surface reconstruction transforms a group of points into a 3D surface. Over the past three decades, the surface reconstruction problem has attracted the attention of many researchers and has been extensively studied. A comprehensive review is given by Berger et al.\cite{BergerTagliasacchi2016}. There exist two classes of surface reconstruction methods, interpolation method and approximation method, where interpolation method can be further categorized into Delaunay method and region growing method. Most of the Delaunay methods \cite{AmentaBern1998, AmentaChoi2000} have been rigorously validated, but require that the input data satisfy the $\epsilon$ sampling condition. Region growing method has been extensively used due to their efficiency and ease of extension from 2D to 3D. These methods start with a seed triangle and expand the reconstructed surface region by advancing the leading edge forward with certain criteria. Wang et al. \cite{WangSu2019} developed a surface reconstruction method based on region growing method and the Delaunay method. A seed triangle is selected from the 3D Delaunay triangulation, and the surface is reconstructed by adding new triangles from the leading edge. The results show that the method compares favorably with popular classical methods when the input points are undirected and the output surfaces are triangles. Thayyil et al. \cite{ThayyilYadav2021} developed a surface reconstruction method. The local Delaunay triangulation is used to accelerate the reconstruction and improve memory efficiency.

In general, interpolation is preferred over approximation to reconstruct the coordinates of points that are used as unique inputs and to preserve more surface detail. However, previous region growing methods required first performing 3D Delaunay triangulation of unorganized input points. But the high complexity of Delaunay triangulation in 3D tends to produce a large number of redundant triangles or tetrahedra, and the Delaunay criterion strictly speaking fails in 3D, where sometimes it is impossible to find a sphere that does not contain any other points anyway. In this study, we developed a novel 3D local surface reconstruction method based on the free-surface particles and the region growing method, and has the following contributions:

(1) Local surface reconstruction is facilitated by the fact that it does not require Delaunay triangulation, but only free-surface particles.

(2) The local surface reconstruction of each particle is carried out independently, which is convenient for parallel processing.

The developed 3D local surface reconstruction method includes selection of the initial edge, principles of triangle expansion, and evaluation function.

\subsubsection{Basic concepts}

\begin{figure}[htbp]
\centering
\includegraphics[width=16cm]{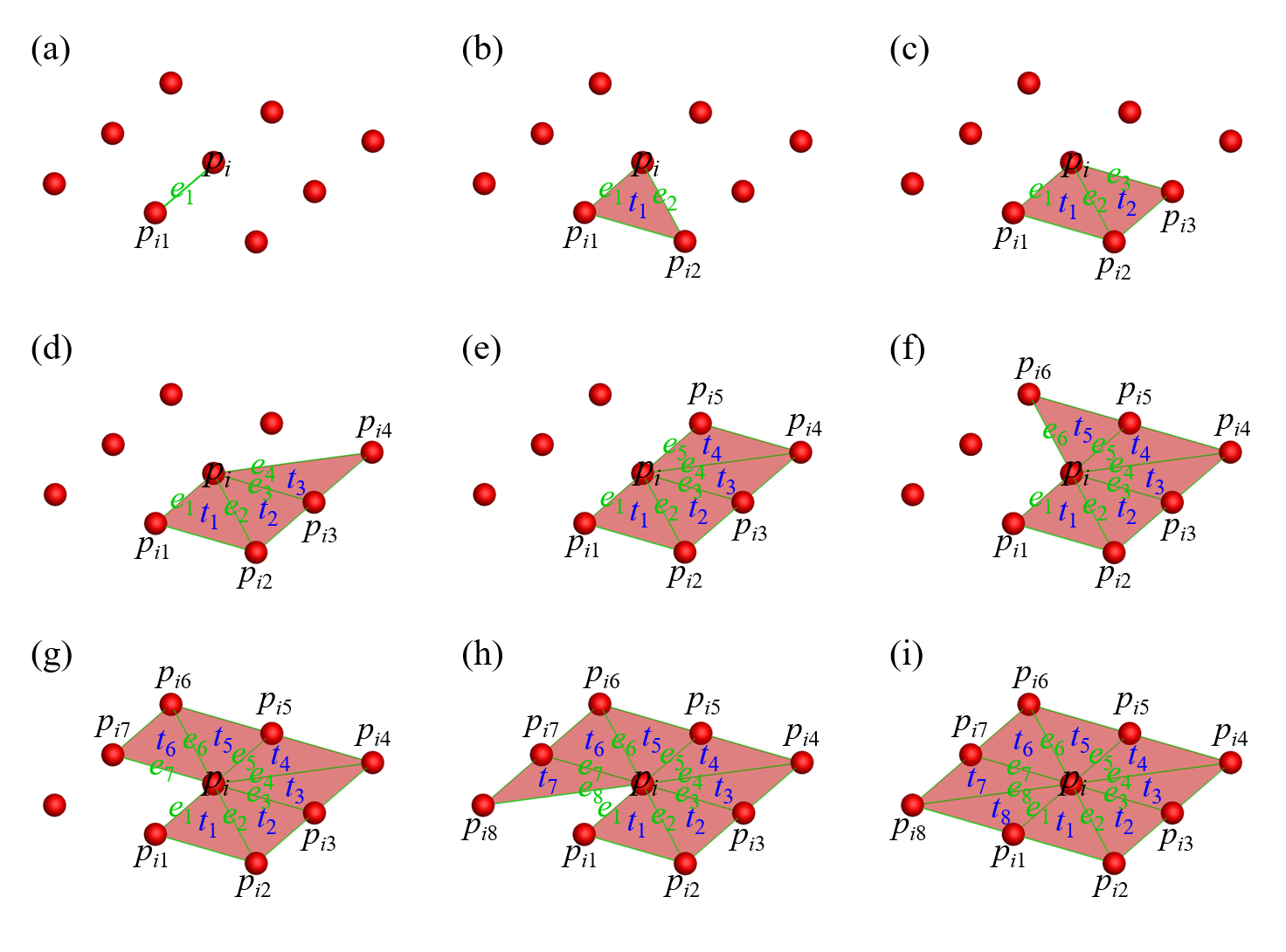}
\captionsetup{font={normalsize,stretch=1.0}}
\caption{The local surface reconstruction process for the free-surface particle $ p_i $. The red particles denote the free-surface particles, the particle $ p_i $ denotes the particle to be reconstructed on the local surface, the particles $ p_{i1} $, $ p_{i2} $, $ \cdots $, $ p_{i8} $ denote the neighboring particles involved in the reconstructed local surface, the edges $ e_1 $, $ e_2 $, $ \cdots $, $ e_8 $ denote the edges involved in the reconstructed local surface, the triangles $ t_1 $, $ t_2 $, $ \cdots $, $ t_8 $ denote the triangles involved in the reconstructed local surface, and $ \left( \rm a \right) $ to $ \left( \rm i \right) $ denotes the advancing process of reconstructing the local surface.}
\label{fig:Figure-3-2-1}
\end{figure}

The local surface reconstruction process for the free-surface particle $ p_i $ is shown in Fig.~\ref{fig:Figure-3-2-1}. The set of neighboring particles involved in the reconstructed local surface is $ \mathbf{P}_s = \left\{ p_{i1}, p_{i2}, \cdots, p_{in}, \notag \right. \\ \left. p_{i\left( n+1 \right)}, p_{i1} \right\} $, where the $ n+1 $ denotes the number of triangles on the reconstructed local surface, and the last particle is the same as the first particle indicating that the reconstructed local surface needs to be closed. The set of edges involved in the reconstructed local surface is $ \mathbf{E}_s = \left\{ e_1, e_2, \cdots, e_n, e_{\left( n+1 \right)}, e_1 \right\} $, where $ e_n $ denotes the edge consisting of particle $ p_i $ and $ p_{in} $, and $ e_1 $ is the initial edge of the reconstructed local surface. The set of triangles involved in the reconstructed local surface is $ \mathbf{T}_s = \left\{t_1, t_2, \cdots, t_n, t_{\left( n+1 \right)} \right\} $, where $ t_n $ is the $n$-th triangle composing the reconstructed local surface and consists of three particles $ p_i $, $ p_{in} $, and $ p_{i(n+1)} $, involved in edges $ e_n $ and $ e_{n+1} $. At each triangle extension, for the edge $ e_n $ to be extended, there exist $ m $ candidate triangles connected to the edge $ e_n $, the set of them is $ \mathbf{T}_{n} = \left\{t_{n\_1}, t_{n\_2}, \cdots, t_{n\_m} \right\} $, and the set of corresponding particles is $ \mathbf{P}_{n} = \left\{p_{i\left( n+1 \right)\_1}, p_{i\left( n+1 \right)\_2}, \cdots, p_{i\left( n+1 \right)\_m} \right\} $.

\subsubsection{Selection of the initial edge}

\begin{figure}[htbp]
\centering
\includegraphics[width=9cm]{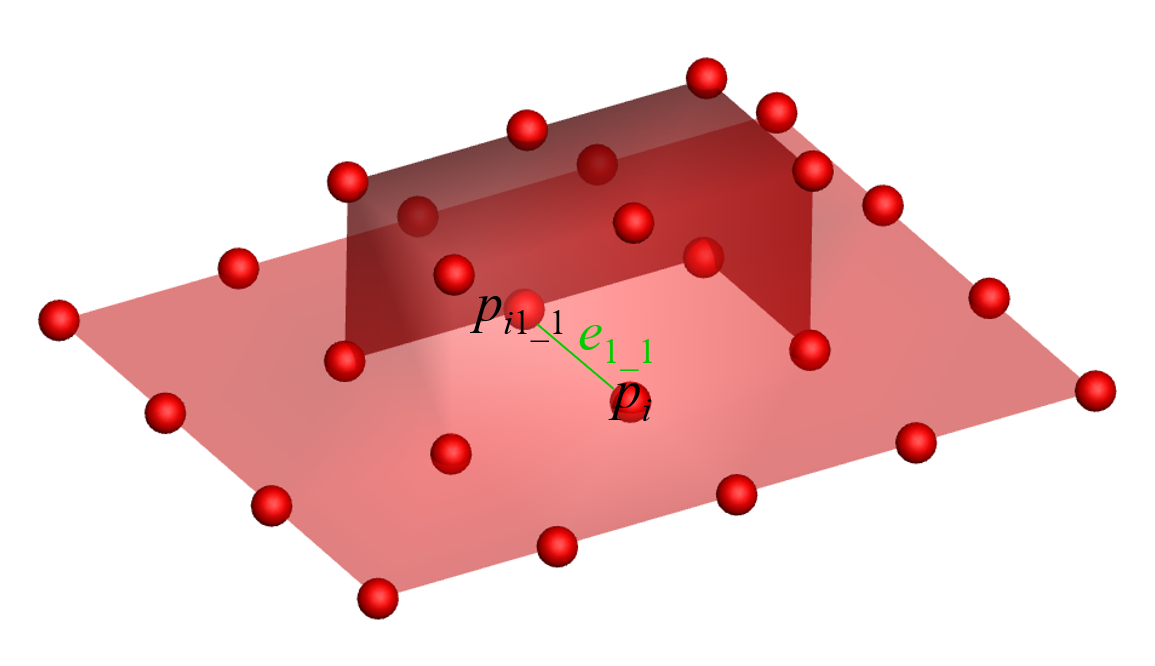}
\captionsetup{font={normalsize,stretch=1.0}}
\caption{Error selection of the initial edge. If the edge $ e_{1\_1} $ consisting of particle $ p_i $ and $ p_{i1\_1} $ is selected as the initial edge, the subsequent triangle expansion will not be able to be conducted according to the principles of triangle expansion because the edge $ e_{1\_1} $ is an inner edge.}
\label{fig:Figure-3-2-2}
\end{figure}

For the free-surface particle $ p_i $ that needs to reconstruct the local surface, the edge consisting of the nearest free-surface particle to the particle $ p_i $ is selected as the initial edge of the reconstructed local surface. However, this selection may have a failure situation, as shown in Fig.~\ref{fig:Figure-3-2-2}. If the edge $ e_{1\_1} $ consisting of particle $ p_i $ and $ p_{i1\_1} $ is selected as the initial edge, the subsequent triangle expansion will not be able to be conducted according to the principles of triangle expansion in Section~\ref{subsubsec:priTriExpansion} because the edge $ e_{1\_1} $ is an inner edge. Then, the initial edge should be replaced, based on the distances of the neighboring free-surface particles from the particle $ p_i $, for several iterations until a triangular expansion can be performed.

\subsubsection{Principles of triangle expansion} \label{subsubsec:priTriExpansion}

To make the reconstructed local surface connected with non-noise particles, the valid surface triangle condition, surface topology condition, and triangle angle condition should to be satisfied in the triangle extension.

\noindent \textbf{Definition 1} (Valid surface triangle condition).

(1) All three particles forming the triangle are free-surface particles;

(2) No other free-surface particles are projected into the triangle along its normal direction;

(3) The positive normal of the triangle points to the outside of the surface and the negative normal of the triangle points to the inside of the surface.

\begin{figure}[htbp]
\centering
\includegraphics[width=8cm]{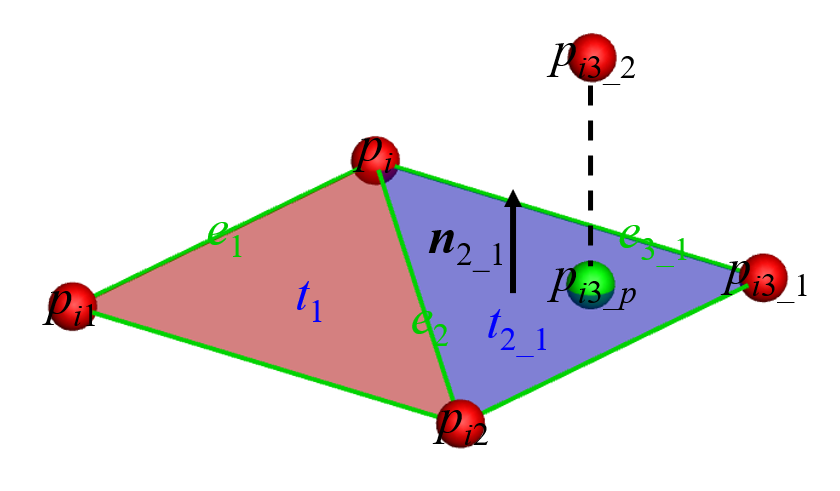}
\captionsetup{font={normalsize,stretch=1.0}}
\caption{Illustration of condition (2) in the valid surface triangle condition.}
\label{fig:Figure-3-2-4}
\end{figure}

\begin{figure}[htbp]
\centering
\includegraphics[width=8cm]{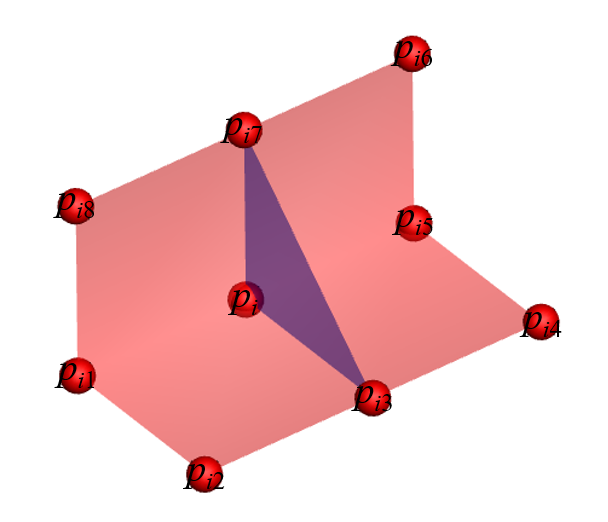}
\captionsetup{font={normalsize,stretch=1.0}}
\caption{Illustration of condition (3) in the valid surface triangle condition.}
\label{fig:Figure-3-2-5}
\end{figure}

In contrast to traditional surface reconstruction methods, where surface triangles are first obtained via the 3D Delaunay triangulation, it is necessary to specify how the surface triangles are valid in the present local surface reconstruction method based on the free-surface particles and the growing method. Condition (1) is used to avoid the introduction of inner particles which leads to topological errors. Condition (2) is used to minimize the introduction of noise particles, as shown in Fig.~\ref{fig:Figure-3-2-4}. For the edge $ e_2 $ to be extended, assuming that there exists a candidate particle $ p_{i3\_1} $, and the particle $ p_{i3\_p} $ projected by $ p_{i3\_2} $ along the normal direction of the triangle $ t_{2\_1} $ is in the interior of the triangle $ t_{2\_1} $. If $ t_{2\_1} $ is added to the local surface of particle $ p_i $ as surface triangle $ t_2 $, the other triangles connected to $ p_{i3\_2} $ can not be added to the surface due to the other triangles with large angles between other triangles and $ t_{2\_1} $ or some other restriction. Condition (3) is also used to prevent surface topology errors, as shown in Fig.~\ref{fig:Figure-3-2-5}. Although the triangle consisting of the particles $ p_i $, $ p_{i3} $, and $ p_{i7} $ meets conditions (1) and (2), the triangle is not a valid surface triangle due to the fact that the upper and lower surfaces of the triangle are pointing towards the outside of the object, and the introduction of the triangle will result in an error in the surface topology of the triangle. Therefore, condition (3) is introduced to further standardize the valid surface triangle.

\noindent \textbf{Definition 2} (Surface topology condition).

(1) An edge can be referenced at most twice.

Condition (1) is used to avoid one edge links with more than 2 surface triangles.

\noindent \textbf{Definition 3} (Triangle angle condition).

(1) The angle between edges $ e_n $ and $ e_{n+1} $ should be greater than $ \theta_{\rm{min}} $ and less than $ \theta_{\rm{max}} $.

(2) The angle between neighboring triangles is greater than $ \theta_s $.

Condition (1) is used to avoid expanding triangles with interior angles that are too large or too small, and the suggested value of $ \theta_{\rm{min}} $ and $ \theta_{\rm{max}} $ are $ 15^{\circ} $ and $ 150^{\circ} $, respectively. Condition (2) is used to avoid a large angle between adjacent triangles \cite{DavidDa2004}, and the suggested value of $ \theta_s $ is $ 60^{\circ} $.

\subsubsection{Evaluation function}

\begin{figure}[htbp]
\centering
\includegraphics[width=7cm]{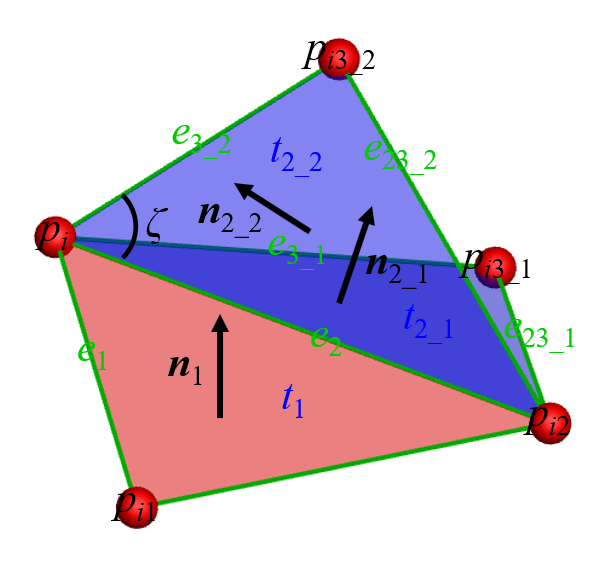}
\captionsetup{font={normalsize,stretch=1.0}}
\caption{Optimal triangle selection in triangle expansion. Red triangles denote surface triangles, blue triangles denote candidate surface triangles.}
\label{fig:Figure-3-2-6}
\end{figure}

After filtering by the principles of triangle expansion in Section~\ref{subsubsec:priTriExpansion}, there is often more than one triangles that match the triangle extension, as shown in Fig.~\ref{fig:Figure-3-2-6}. There are always three requirements to consider in the growth of a triangle: smoothness, triangle interior angle, and small triangle. However, it is difficult to find a triangle that can satisfy the above requirements at the same time. Therefore, an  evaluation function is introduced to balance the three requirements of smoothness, triangle interior angle, and small triangle, as shown as follows:
\begin{equation}
  f
  =
  \xi_1 \left( 1 - \cos \psi \right)
  +
  \xi_2 \left\| \cos \frac{\pi}{6} - \cos \zeta \right\|
  +
  \xi_3 \frac{\left| e_{23} \right| + \left| e_3 \right|}{\left| e_2 \right|},
\end{equation}
In the above formula, $ \cos \psi $ is the cosine of the angle between the normal vectors of triangles $ t_1 $ and $ t_{2\_m} $. It can be seen that the smaller $ 1 - \cos \psi $ is, the closer the angle between the triangles $ t_1 $ and $ t_{2\_m} $ is to $ 180^{\circ} $ and the smoother the surface is. $ \cos \zeta $  is the cosine of the angle between edges $ e_2 $ and $ e_{3\_m} $, and it is usually assumed that the triangle has the best quality when $ \zeta $ is equal to $ 60^{\circ} $. $ \left| e_{23} \right| + \left| e_3 \right| $ denotes the sum of the two sides of the new triangle, the smaller the value, the smaller the newly generated triangle on the surface. $ \xi_1 $, $ \xi_2 $ and $ \xi_3 $ are weight coefficients of smoothness, triangle interior angle, and small triangle, and $ \xi_1 + \xi_2 + \xi_3 = 1 $.

\subsection{Detection of surface-surface contact}

The 3D solution domain is divided into cubes with equal sides, and all particles are placed inside the cubes. For each particle, all particles within its cube and its surroundings (total 27), including particles belonging to the same and different objects, are searched and labeled. This process will be performed at the beginning of the SPH solution and is also effective for solving the governing equations.

\begin{figure}[htbp]
\centering
\includegraphics[width=10cm]{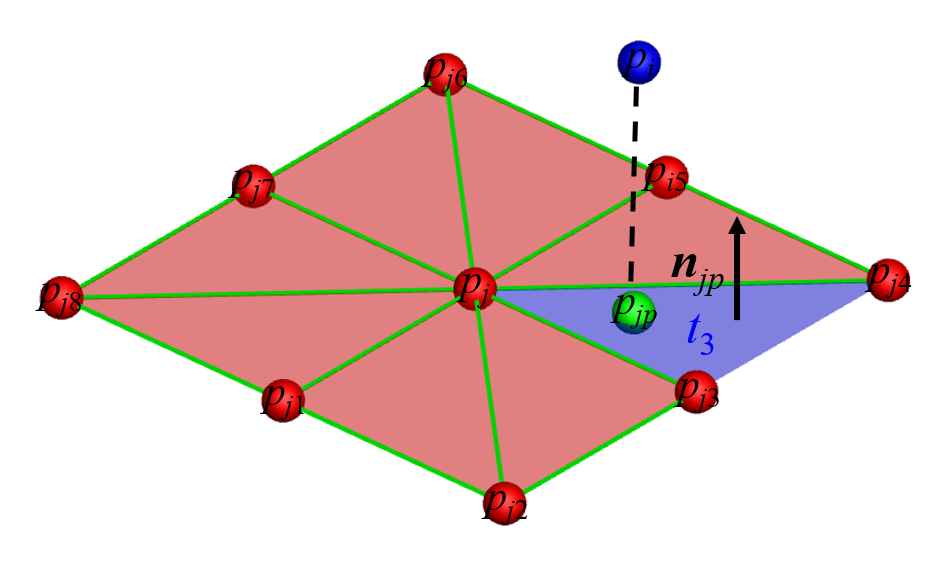}
\captionsetup{font={normalsize,stretch=1.0}}
\caption{Schematic of the detection of surface-surface contact. The blue particle $ p_i $ denotes the slave particle, the red particle $ p_j $ denotes the master particle, the red particles $ p_{j1}, p_{j2}, \cdots, p_{j8} $ denote the particles involved in the reconstructed local surface of the particle $ p_j $, the particle $ p_{jp} $ denotes the projected particle from the particle $ p_i $ on the master surface, and the $ \boldsymbol{n}_{jp} $ is the unit outward normal vector of the projected triangle of the master surface.}
\label{fig:Figure-3-3-1}
\end{figure}

For each slave particle, search for the nearest particle in its neighborhood that belongs to the master object and project it into each triangle of the local surface of the master particle, check whether the projected particle is inside the triangle, and if there exists a projected particle that is inside the triangle, create slave particle-master surface contact pair and mark the projected particle as $ p_{jp} $, as shown in Fig.~\ref{fig:Figure-3-3-1}. For the slave particle $ p_i $,  its penetration distance to the master surface $ d_{ij}^p $ is calculated as
\begin{equation}
  d_{ij}^p = \left( \boldsymbol{x}_{i}^{p} - \boldsymbol{x}_{j}^{p} \right) \cdot \boldsymbol{n}_{jp},
\end{equation}
where $ \boldsymbol{x}^{p} $ denotes the predicted position without considering the contact forces, and $ \boldsymbol{n}_{jp} $ is the unit outward normal vector of the projected triangle of the master surface. The predicted position of the projected particle is
\begin{equation}
  \boldsymbol{x}_{jp}^{p} = \boldsymbol{x}_{i}^{p} - d_{ij}^p \boldsymbol{n}_{jp}.
\end{equation}

If the penetration distance $ d_{ij}^p $ is less than the initial particle distance, it is determined that the contact occurs for the slave particle-master surface contact pair in the current time step.

\subsection{Enforcement of surface-surface contact}

When the surface-surface contact occurs, the rate of penetration between the slave particle-master surface contact pair is:
\begin{equation}\label{eq:penRate}
  \dot{d}_{n}^{p}
  =
  \left( \boldsymbol{v}_{i}^{p} - \boldsymbol{v}_{jp}^{p} \right) \cdot \boldsymbol{n}_{jp},
\end{equation}
where the predicted velocity of the projected particle $ \boldsymbol{v}_{jp}^{p} $ is obtained by interpolating the velocities of the three particles of the projected triangle on the master surface. Normal penalty force is applied only when the rate of penetration is less than zero.

The velocity should be modified by applying a normal penalty force $ \boldsymbol{f}_{ij}^{n} $ to make sure that the penetration between the slave particle-master surface contact pair is zero at the end of the current computation step. Since the kick-drift-kick scheme is employed in this study, the modified velocity obtained using the normal penalty force $ \boldsymbol{f}_{ij}^{n} $ is
\begin{equation}\label{eq:modVelInPari}
  \boldsymbol{v}_{i}^{c}
  =
  \boldsymbol{v}_{i}^{p} + \frac{\boldsymbol{f}_{ij}^{n}}{m_i} \frac{\Delta t}{2},
\end{equation}
\begin{equation}\label{eq:modVelInParjc}
  \boldsymbol{v}_{jp}^{c}
  =
  \boldsymbol{v}_{jp}^{p} - \frac{\boldsymbol{f}_{ij}^{n}}{m_{jp}} \frac{\Delta t}{2}.
\end{equation}

Applying the impermeability conditions leads to
\begin{equation}\label{eq:abc}
  \dot{d}_{n}^{c}
  =
  \left( \boldsymbol{v}_{i}^{c} - \boldsymbol{v}_{jp}^{c} \right) \cdot \boldsymbol{n}_{jp} = 0.
\end{equation}

Substituting Eqs.~\refeq{eq:modVelInPari} and ~\refeq{eq:modVelInParjc} into the Eq.~\refeq{eq:abc} and solving for the normal penalty force:
\begin{equation}\label{eq:norPenForce}
  \boldsymbol{f}_{ij}^{n}
  =
  \displaystyle \frac{\left[ \left( \boldsymbol{v}_{i}^{p} - \boldsymbol{v}_{jp}^{p} \right) \cdot \boldsymbol{n}_{jp} \right] \boldsymbol{n}_{jp}}{\displaystyle \frac{1}{2}\Delta tm_i + \displaystyle \frac{1}{2} \Delta tm_{jp}}
\end{equation}

Regarding the tangential friction force $ \boldsymbol{f}_{ij}^{\tau} $, we first assume a state of static friction
\begin{equation}
  \dot{d}_{\tau}
  =
  \left\| \left( \boldsymbol{v}_{i}^{p}-\boldsymbol{v}_{jp}^{p} \right) -\left[ \left( \boldsymbol{v}_{i}^{p}-\boldsymbol{v}_{jp}^{p} \right) \boldsymbol{n}_{jp} \right] \boldsymbol{n}_{jp} \right\| =0
\end{equation}
where $ \dot{d}_{\tau} $ denotes the rate of relative sliding. The tangential friction force $ \boldsymbol{f}_{ij}^{\tau ,sticking} $ is obtained as:
\begin{equation}\label{eq:tanFriForStatic}
  \boldsymbol{f}_{ij}^{\tau ,sticking}
  =
  \displaystyle \frac{\left( \boldsymbol{v}_{i}^{p}-\boldsymbol{v}_{jp}^{p} \right) -\left[ \left( \boldsymbol{v}_{i}^{p}-\boldsymbol{v}_{jp}^{p} \right) \boldsymbol{n}_{jp} \right] \boldsymbol{n}_{jp}}{\displaystyle \frac{1}{2}\Delta tm_i+ \displaystyle \frac{1}{2}\Delta tm_{jp}}
\end{equation}
The applied tangential friction force $ \boldsymbol{f}_{ij}^{\tau ,sticking} $ makes sure that there is no relative sliding of the slave particle-master surface contact pair at the end of the current computational step.

When the static friction force exceeds the upper limit, the friction between the slave particle-master surface contact pair will change from static friction to dynamic friction and relative sliding will occur. The tangential friction force considering both static and dynamic friction is calculated as follows
\begin{equation}\label{eq:tanFriForce}
  \boldsymbol{f}
  =
  \left\{
  \begin{array}{lll}
    \mu \left\| \boldsymbol{f}_{ij}^{n} \right\| \frac{\boldsymbol{f}_{ij}^{\tau ,sticking}}{\left\| \boldsymbol{f}_{ij}^{\tau ,sticking} \right\|}
    &
    \left\| \boldsymbol{f}_{ij}^{\tau ,sticking} \right\| \geqslant \mu \left\| \boldsymbol{f}_{ij}^{n} \right\| \medskip \\
    \boldsymbol{f}_{ij}^{\tau ,sticking}
    &
    \left\| \boldsymbol{f}_{ij}^{\tau ,sticking} \right\| < \mu \left\| \boldsymbol{f}_{ij}^{n} \right\|
  \end{array}
  \right.
\end{equation}
where $ \mu $ is the frictional coefficient between the contact pair.

To summarize, the contact force applied to the slave particle is obtained from Eqs.~\refeq{eq:norPenForce} and ~\refeq{eq:tanFriForce}, and then added to the Eq.~\refeq{eq:velocity-half-step}. The contact force applied to the three particles of the projected triangle on the master surface can be solved by Newton's third law \cite{XiaoLiu20232, DongHao2023}.

\subsection{Supplementary particle-particle contact}

\begin{figure}[htbp]
\centering
\includegraphics[width=12cm]{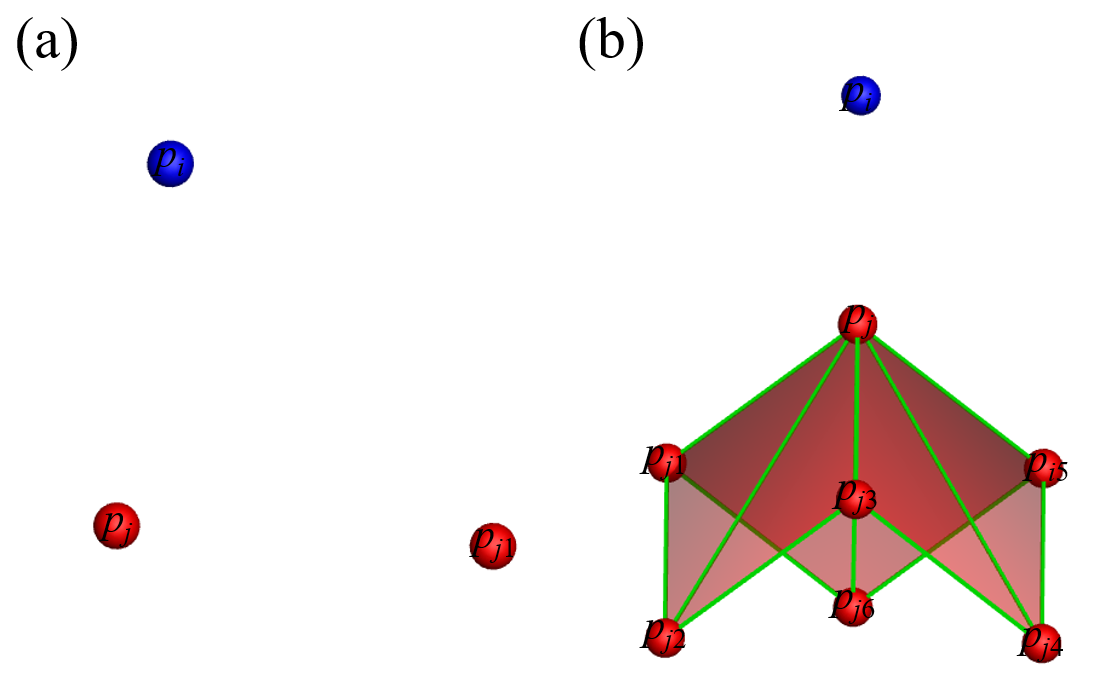}
\captionsetup{font={normalsize,stretch=1.0}}
\caption{The cases of surface-surface contact failure. (a) The inability to reconstruct the local surface of the master particle $ p_j $ when the object undergoes large deformation; (b) The slave particle $ p_i $ may not be able to project on the local surface of the master particle when the master particle $ p_j $ is a convex particle.}
\label{fig:Figure-3-5-1}
\end{figure}

Despite the high accuracy and ease of the surface-surface contact method, however, it will fail in some cases, such as the inability to reconstruct the local surface when the object undergoes large deformation, and the slave particle may not be able to project on the local surface of the master particle when the master particle is a convex particle, as shown in Fig.~\ref{fig:Figure-3-5-1}. Therefore, the particle-particle contact method is supplemented in 3D hybrid contact method to solve the above problems. In the 3D particle-particle contact method, only the calculation of the normal penalty force is performed, where the direction of the penalty force is calculated as follows.
\begin{equation}
  \boldsymbol{n}_j
  =
  \frac{\boldsymbol{x}_i - \boldsymbol{x}_j}{\left\| \boldsymbol{x}_i - \boldsymbol{x}_j \right\|}
\end{equation}
Then it is brought to Eq.~\refeq{eq:penRate} to calculate the rate of penetration and the normal penalty force.

\section{Results and discussion}\label{sec:4}

In this section, three numerical cases accompanied by strong interface relative motion are presented to validate the 3D hybrid contact method proposed in this study. First, numerical simulations of a block sliding down a slope are carried out to investigate the velocity and displacement of the block in the frictionless and friction cases, respectively, and the results are compared with the analytical solution and the particle-particle contact method \cite{CampbellVignjevic2000}. Then, the SPH method is combined with a viscoelastic-viscoplastic-damage constitutive model and a hot spot model to study the ignition response of high explosives under crack extrusion loading, and the results are compared with the traditional momentum equation method \cite{LiberskyPetschek1993} and the particle-particle contact method \cite{CampbellVignjevic2000}. Finally, numerical simulations of the high velocity impact are carried out, focusing on the detection accuracy and detection time of the improved high accuracy free-surface particle detection method, the results of the local surface reconstruction method, and the applicability of the 3D hybrid contact method in the extreme deformation case.

\subsection{A block sliding along a slope}

\begin{figure}[htbp]
\centering
\includegraphics[width=16cm]{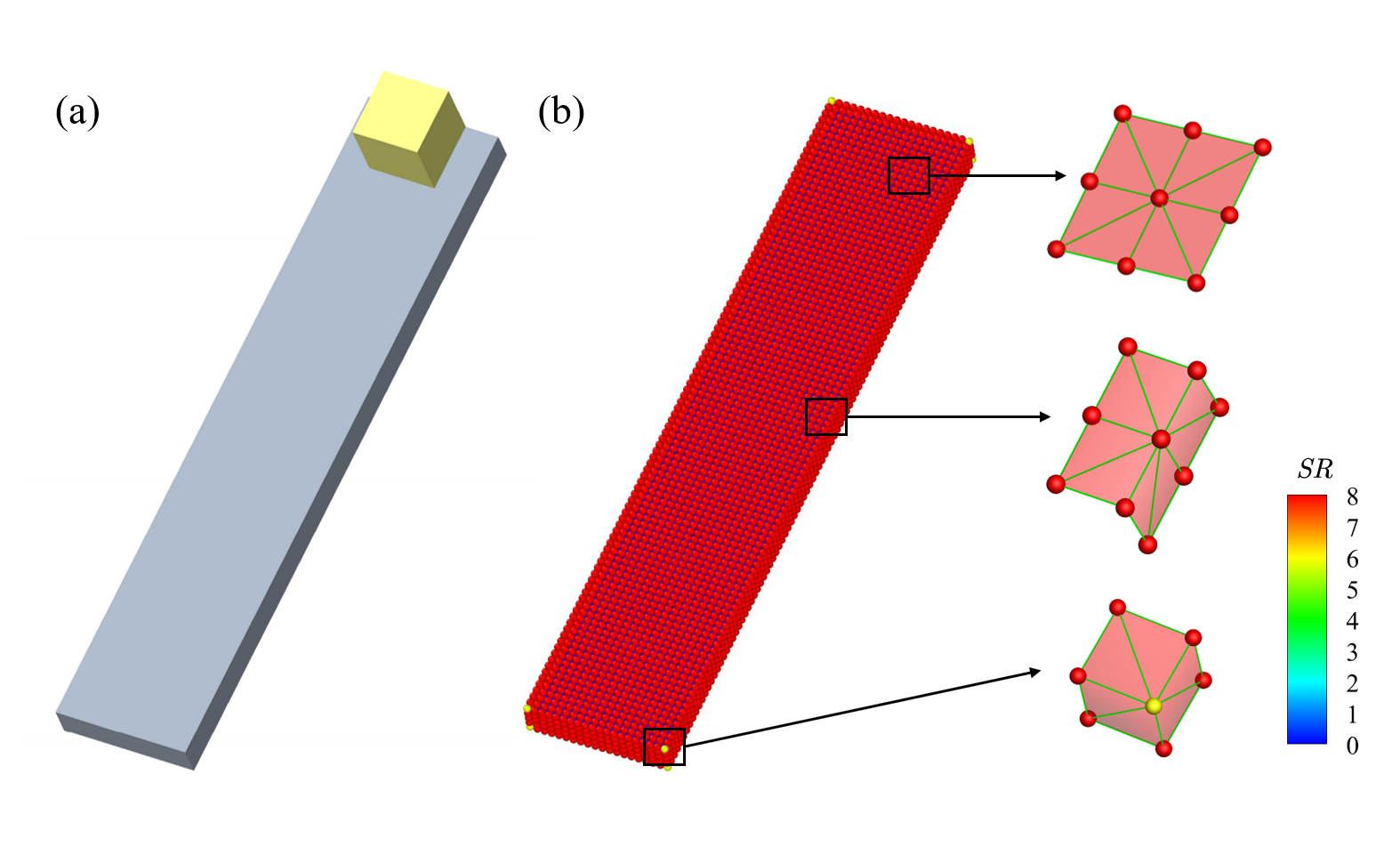}
\captionsetup{font={normalsize,stretch=1.0}}
\caption{(a) Computational model of a block sliding under gravity along a slope; (b) The result of the local surface reconstruction of the slope. Since the local surface reconstruction of different free-surface particles is independent of each other and their surface triangles overlap, the overall reconstructed surface cannot be given, but the number of surface triangles $ SR $ for free-surface particles and the reconstructed local surfaces of some characteristic particles are given.}
\label{fig:Figure-4-1-1}
\end{figure}

In this subsection, a simulation of a block sliding under gravity along a slope with an inclination of $ 30^{\circ} $ is carried out, and the computational model is shown in Fig.~\ref{fig:Figure-4-1-1}(a). The block is a cube with sides of 1 m. The slope is a plate 10 m long, 2.0 m wide, and 0.5 m thick. A elastic constitutive model is used, with the same material parameters for the block and the slope. The density is 7850 $ \mathrm{kg}/\mathrm{m}^3 $, Young's modulus $ E $ is 210 GPa, and Poisson's ratio $ \nu $ is 0.3. The gravitational acceleration is set to 9.8 $ \mathrm{m}/\mathrm{s}^2 $ in the $ z $ direction. The initial particle spacing is 0.1 mm. In the simulations, the slope surface is selected as the master surface. Since there is no large deformation on the slope surface, the local surfaces of all particles of the slope are reconstructed initially, and subsequent calculations are not required. The result of the local surface reconstruction of the slope is shown in Fig.~\ref{fig:Figure-4-1-1}(b).

\begin{figure}[htbp]
\centering
\includegraphics[width=16cm]{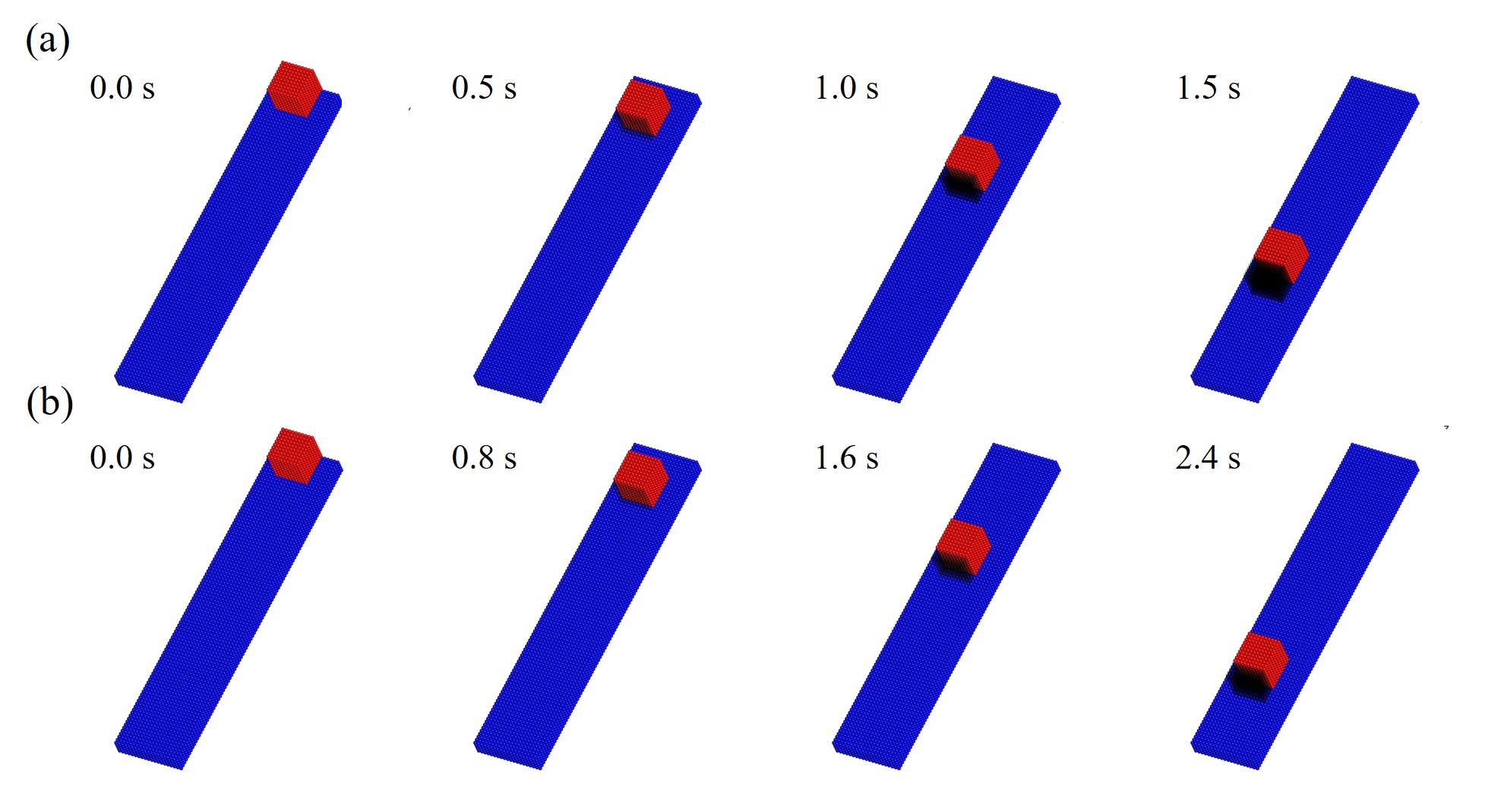}
\captionsetup{font={normalsize,stretch=1.0}}
\caption{Calculation results of the block sliding under gravity along a slope. (a) $ \mu $ = 0; (b) $ \mu $ = 0.3.}
\label{fig:Figure-4-1-2}
\end{figure}

\begin{figure}[htbp]
\centering
\includegraphics[width=16cm]{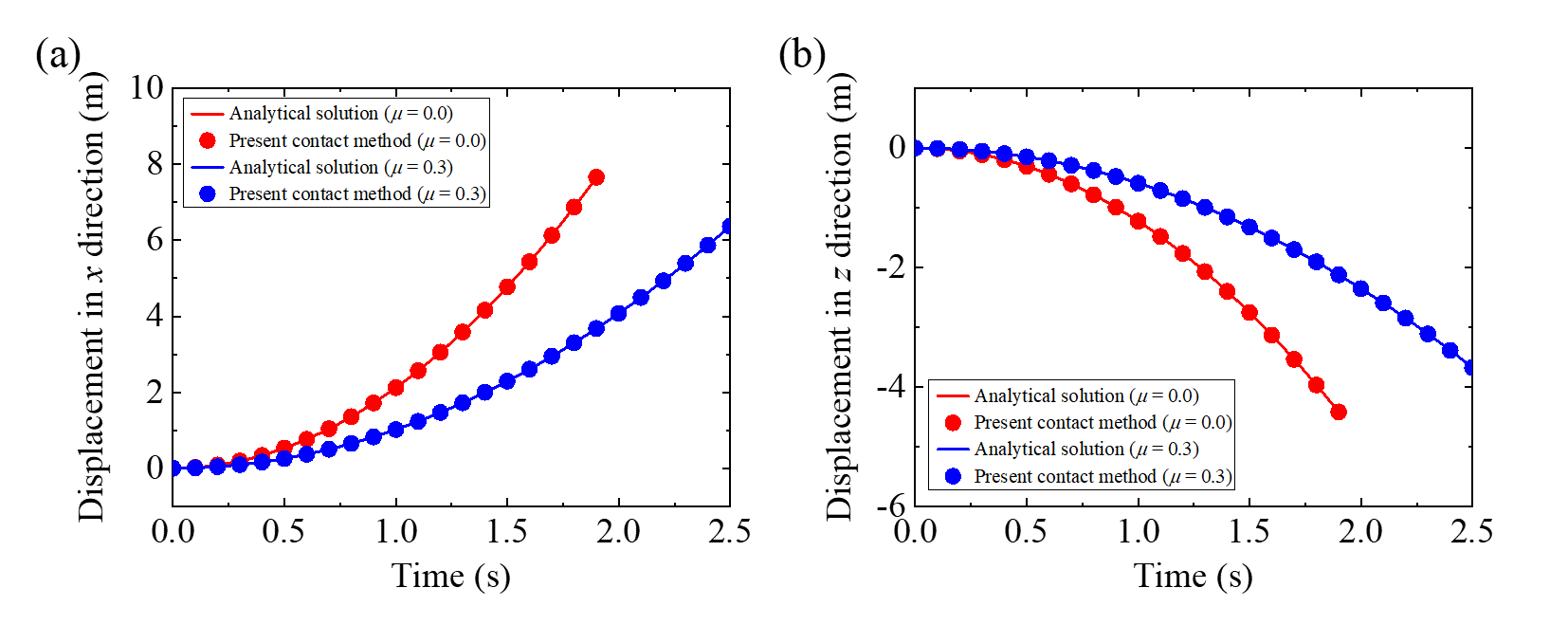}
\captionsetup{font={normalsize,stretch=1.0}}
\caption{The displacements of the block in the $ x $ and $ z $ directions.}
\label{fig:Figure-4-1-3}
\end{figure}

Figure~\ref{fig:Figure-4-1-2} presents the block sliding process in both frictionless and friction situations, where the arrows represent the velocity vectors of the block particles. The block velocity decreases as the coefficient of dynamic friction increases, and the velocity vectors of all the block particles are equal in magnitude and move in a direction parallel to the surface of the slope during the sliding process. The displacements of the block in the $ x $ and $ z $ directions are further given in Fig.~\ref{fig:Figure-4-1-3} and compared with the analytical solution, and the simulation results agree with the analytical solution, which demonstrates that the present contact method has high accuracy in dealing with small deformation problems.

\begin{figure}[htbp]
\centering
\includegraphics[width=16cm]{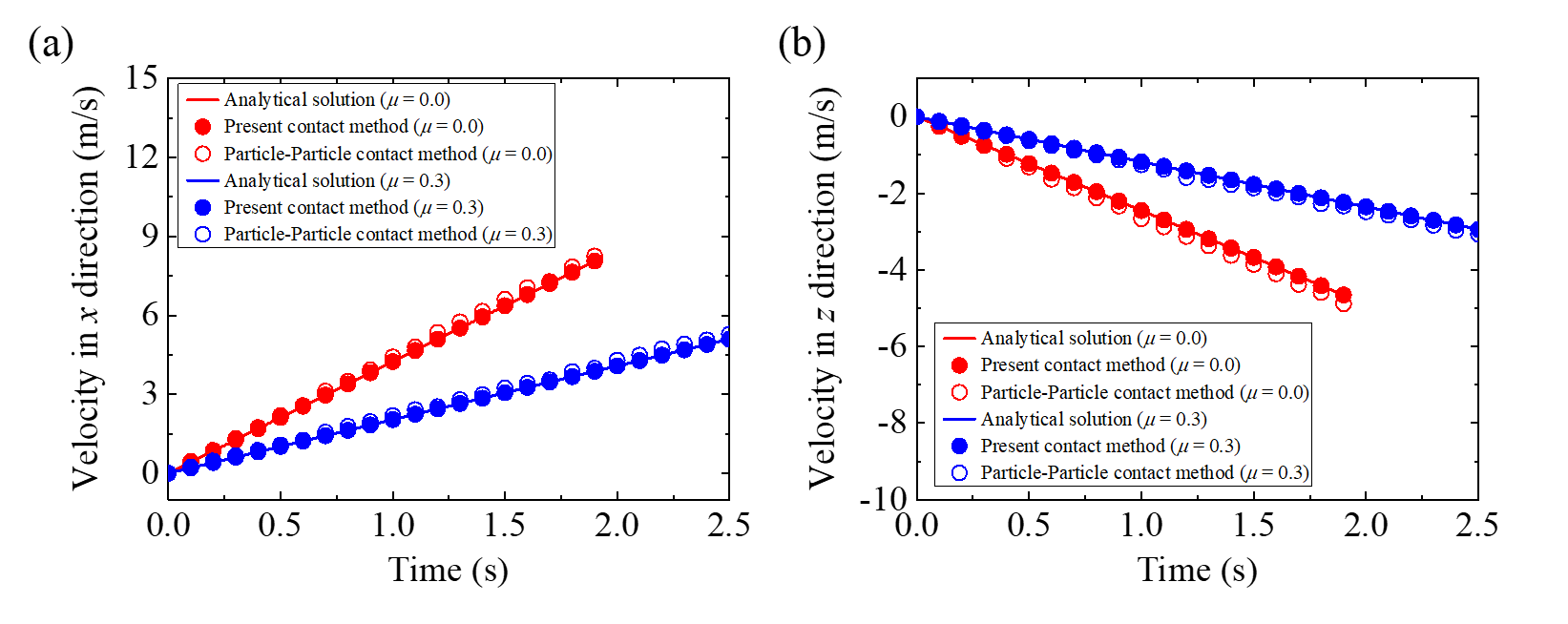}
\captionsetup{font={normalsize,stretch=1.0}}
\caption{The velocity of the block in the $ x $ and $ z $ directions.}
\label{fig:Figure-4-1-4}
\end{figure}

In addition, the velocity of the block in the $ x $ and $ z $ directions are given in Fig.~\ref{fig:Figure-4-1-4}, and the particle-particle contact method \cite{CampbellVignjevic2000} is employed for comparison. The results show that the particle-particle contact method suffers from velocity oscillations due to lower accuracy, while the present contact method does not suffer from velocity oscillations and the results agree with the analytical solution, which further proves the high accuracy of the present contact method in dealing with small deformation problems.

\subsection{Ignition response of high explosives under crack extrusion loading}

High explosives are extensively applied in defense and aerospace applications such as munitions and rocked propellants. During operation, assembly, transportation, etc., high explosives are inevitably subjected to accidental mechanical/thermal excitation, such as falling, impact, fire, or prolonged exposure to high temperatures. These may lead to ignition, which results in varying degrees of energy output, such as combustion, explosion, or even detonation. Due to the coupling of force-heat-chemical reactions during ignition, there is currently a lack of means to accurately predict their ignition safety. The study of the ignition response of high explosives is of great significance in evaluating the safety of munitions and guiding the design of munitions \cite{Asay2010, Leiber2003}.

\begin{figure}[htbp]
\centering
\includegraphics[width=3cm]{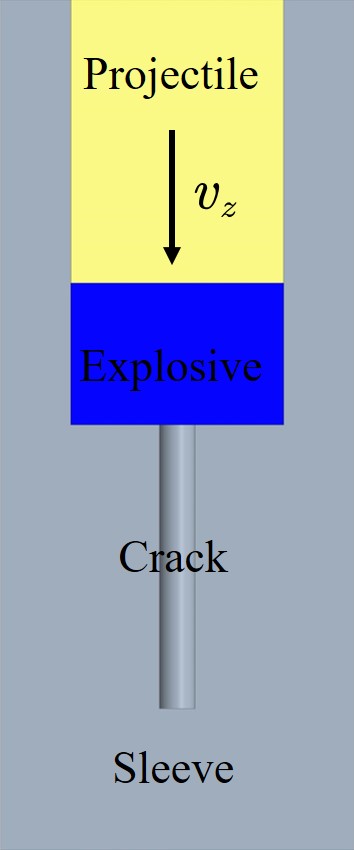}
\captionsetup{font={normalsize,stretch=1.0}}
\caption{Computational model of high explosives under crack extrusion loading.}
\label{fig:Figure-4-2-1}
\end{figure}

\begin{figure}[htbp]
\centering
\includegraphics[width=16cm]{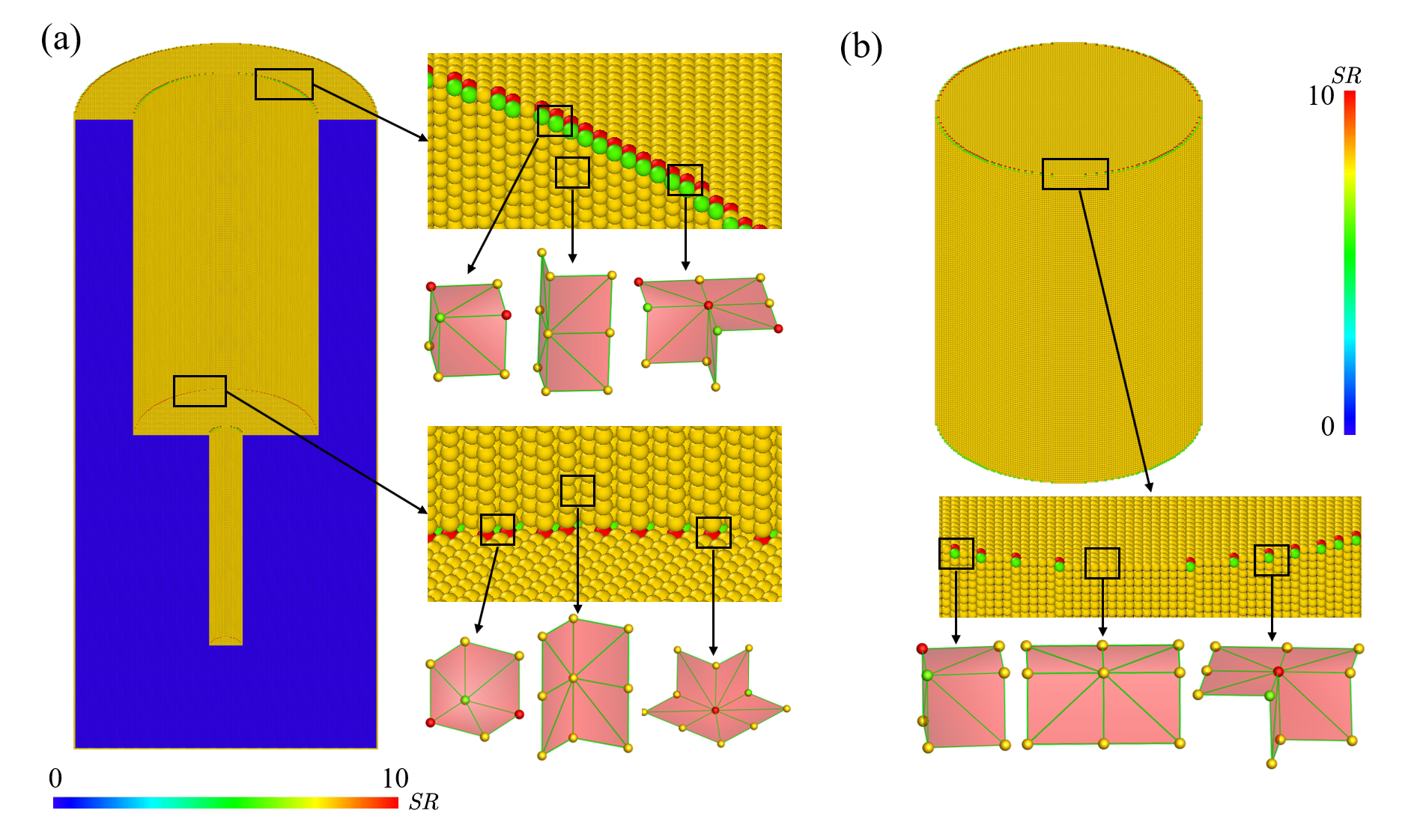}
\captionsetup{font={normalsize,stretch=1.0}}
\caption{Results of reconstructed local surface of free-surface particles on the sleeve and projectile surfaces. Since the local surface reconstruction of different free-surface particles is independent of each other and their surface triangles overlap, the overall reconstructed surface cannot be given, but the number of surface triangles $ SR $ for free-surface particles and the reconstructed local surfaces of some feature particles are given. (a) Sleeve; (b) Projectile.}
\label{fig:Figure-4-2-2}
\end{figure}

Figure~\ref{fig:Figure-4-2-1} gives the computational model of high explosives under crack extrusion loading. The high explosives are placed in a sleeve and a projectile is used to impact the high explosives. There is a cylindrical crack below the high explosives, which flows into the crack and is crushed for rapid flow. The high explosives is a cylinder 30 mm in diameter and 20 mm in height, made of PBX 9501. The projectile is a cylinder 30 mm in diameter and 40 mm in height. The cylindrical sleeve has a lateral thickness of 10 mm at the top and 22.5 mm at the bottom. Both the projectile and the sleeve are made of steel. The PBX 9501 is modeled with the Mie-Gr$\rm {\ddot{u}}$neisen equation of state the viscoelastic-viscoplastic-damage constitutive model \cite{LiuHuang2019, LiuHuang2020}, and a mesoscopic model based on frictional heat generation from a microcrack is introduced as a hot spot model \cite{BennettHaberman1998}.  The steel is modeled with the Mie-Gr$\rm {\ddot{u}}$neisen equation of state and the Johnson-Cook constitutive model. The material parameters of PBX 9501 and steel are shown in Tables~\ref{tab:Table-4-2-1},~\ref{tab:Table-4-2-2}, and~\ref{tab:Table-4-2-3}, respectively. The initial impact velocity of the projectile is -200 m/s in the $ z $ direction. The initial particle distance is 0.25 mm and the total number of particles is 15034880. In the simulations, when the contact between the high explosives and the sleeve occurs, the sleeve surface is selected as the master surface. When the contact between the high explosives and the projectile occurs, the projectile surface is selected as the master surface. When the contact between the projectile and the sleeve occurs, the sleeve surface is selected as the master surface. Since there is no large deformation on the sleeve and projectile surface, the local surfaces of all particles of the sleeve and projectile are reconstructed initially, and subsequent calculations are not required. The results of the local surface reconstruction of the sleeve and projectile are shown in Fig.~\ref{fig:Figure-4-2-2}. The friction force on the material surfaces is not considered in the calculations.

\begin{table}[htbp]
\centering
\caption{Constitutive model parameters for PBX 9501}
\begin{tabular}
{p{23mm}<{\centering}
 p{23mm}<{\centering}
 p{23mm}<{\centering}
 p{23mm}<{\centering}
 p{23mm}<{\centering}
 p{23mm}<{\centering}
}
\toprule

 $ \rho_0 \left( \rm kg / cm^3 \right) $ &
 $ \nu $ &
 $ \tau^{\left( 1 \right)} \left( \upmu \rm s \right) $ &
 $ \tau^{\left( 2 \right)} \left( \upmu \rm s \right) $ &
 $ \tau^{\left( 3 \right)} \left( \upmu \rm s \right) $ &
 $ \tau^{\left( 4 \right)} \left( \upmu \rm s \right) $ \\

\midrule

 1840.0 &
 0.4 &
 10000.0 &
 1000.0 &
 100.0 &
 10.0 \\

\midrule

 $ G_{\infty}\left( \mathrm{MPa} \right) $ &
 $ G^{\left( 1 \right)}\left( \mathrm{MPa} \right) $ &
 $ G^{\left( 2 \right)}\left( \mathrm{MPa} \right) $ &
 $ G^{\left( 3 \right)}\left( \mathrm{MPa} \right) $ &
 $ G^{\left( 4 \right)}\left( \mathrm{MPa} \right) $ &
 \\

\midrule

 1946.79 &
 1413.12 &
 491.798 &
 934.135 &
 1566.39 &
 \\

\midrule

 $ Z_0\left( \mathrm{MPa} \right) $ &
 $ Z_1\left( \mathrm{MPa} \right) $ &
 $ \beta_0 $ &
 $ \beta_1 $ &
 $ m_0 $ &
 $ m_1 $ \\

\midrule

 363.51 &
 868.56 &
 0.0 &
 5.39 &
 6.27 &
 0.2 \\

\midrule

 $ D_0\left( {\upmu \rm s}^{-1} \right) $ &
 $ h $ &
  &
  &
  &
 \\

\midrule

 100.0 &
 0.65 &
  &
  &
  &
 \\

\midrule

 $ a\left( \mathrm{mm} \right) $ &
 $ c_0\left( \mathrm{mm} \right) $ &
 $ v_{\max}\left( \mathrm{km}/\mathrm{s} \right) $ &
 $ K_0\left( \mathrm{MPa}\cdot \mathrm{s}^{\frac{1}{2}} \right) $ &
 $ m $ &
 $ \mu _c $ \\

\midrule

 0.025 &
 0.003 &
 0.3 &
 2.0 &
 6 &
 0.55 \\

\midrule

 $ \mu_t $ &
 $ v_a $ &
 $ v_b $ &
 $ \dot{\bar{\varepsilon}}_0 \left( {\upmu \rm s}^{-1} \right) $ &
  &
 \\

\midrule

 1.22 &
 0.89 &
 -1.06 &
 0.00001 &
  &
 \\

\bottomrule
\end{tabular}
\label{tab:Table-4-2-1}
\end{table}

\begin{table}[htbp]
\centering
\caption{Hot spot model parameters for PBX 9501}
\begin{tabular}
{p{28mm}<{\centering}
 p{28mm}<{\centering}
 p{28mm}<{\centering}
 p{28mm}<{\centering}
 p{28mm}<{\centering}
}
\toprule

 $ k\left( \mathrm{W}/\left( \mathrm{m}\cdot \mathrm{K} \right) \right) $ &
 $ c\left( \mathrm{J}/\left( \mathrm{kg}\cdot \mathrm{K} \right) \right) $ &
 $ \Delta H\left( \mathrm{J}/\mathrm{Kg} \right) $ &
 $ Z\left( \displaystyle \frac{1}{\mathrm{s}} \right) $ &
 $ E/R\left( \mathrm{K} \right) $ \\

\midrule

 0.5 &
 1200.0 &
 $ 5.5\times 10^6 $ &
 $ 5.0\times 10^{19} $ &
 26520.0 \\

\bottomrule
\end{tabular}
\label{tab:Table-4-2-2}
\end{table}

\begin{table}[htbp]
\centering
\caption{Johnson-Cook model parameters for steel}
\begin{tabular}
{p{23mm}<{\centering}
 p{23mm}<{\centering}
 p{23mm}<{\centering}
 p{23mm}<{\centering}
 p{23mm}<{\centering}
 p{23mm}<{\centering}
}
\toprule

 $ \rho_0 \left( \rm kg / cm^3 \right) $ &
 $ E\left( \mathrm{MPa} \right) $ &
 $ \nu $ &
 $ A\left( \mathrm{MPa} \right) $ &
 $ B\left( \mathrm{MPa} \right) $ &
 $ n $ \\

\midrule

 7850.0 &
 212000.0 &
 0.26 &
 249.2 &
 45.6 &
 0.875 \\

\midrule

 $ c $ &
 $ m $ &
  &
  &
  &
 \\

\midrule

 0.32 &
 0.76 &
  &
  &
  &
 \\

\bottomrule
\end{tabular}
\label{tab:Table-4-2-3}
\end{table}

\begin{figure}[htbp]
\centering
\includegraphics[width=16cm]{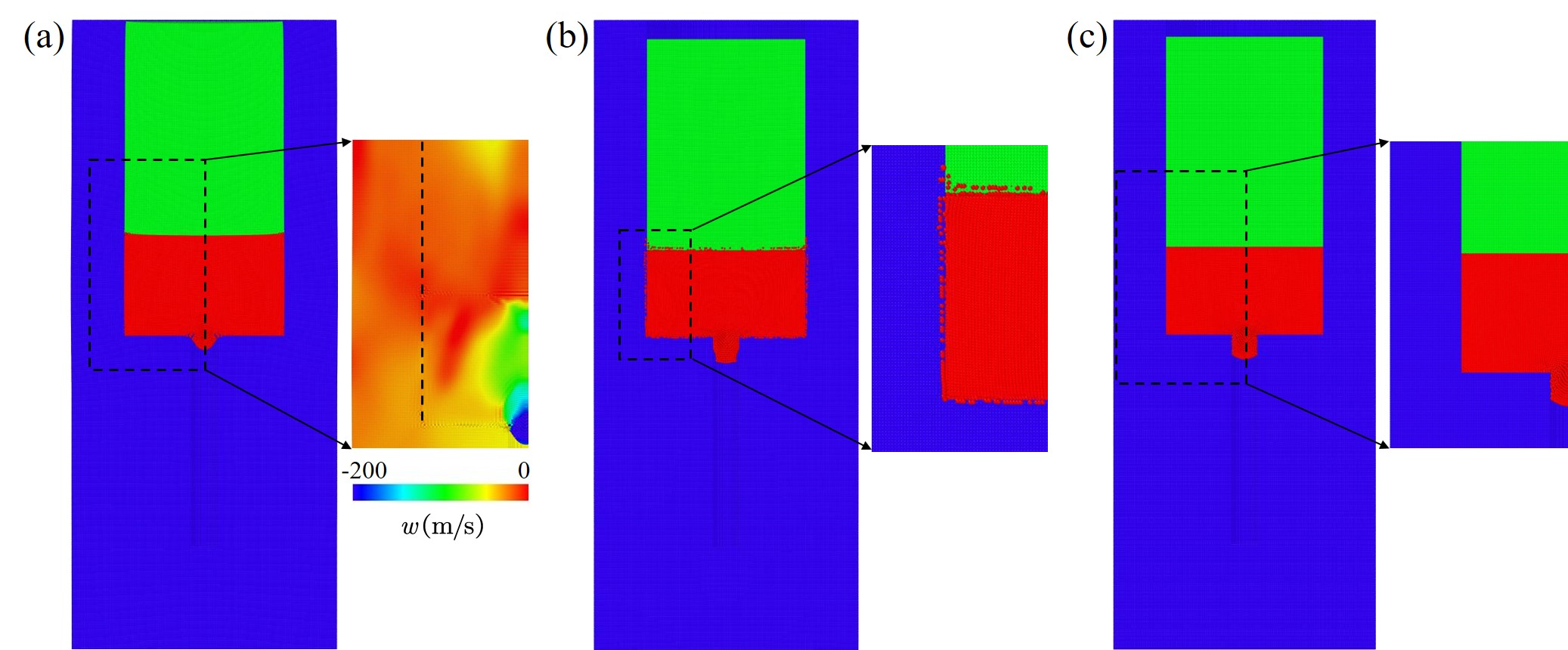}
\captionsetup{font={normalsize,stretch=1.0}}
\caption{The simulation results of high explosives under crack extrusion loading with different contact methods. (a) Momentum equation method \cite{LiberskyPetschek1993}; (b) Particle-particle contact method \cite{CampbellVignjevic2000}; (c) Present contact method.}
\label{fig:Figure-4-2-3}
\end{figure}

Figure~\ref{fig:Figure-4-2-3} presents the simulation results of high explosives under crack extrusion loading with different contact methods. For the momentum equation method \cite{LiberskyPetschek1993}, even though there is no friction force between the projectile and the sleeve, the velocity of the projectile rapidly decreases to zero due to the introduction of virtual shear and tensile stresses in the momentum equation, which does not follow the theoretical situation. For the particle-particle contact method \cite{CampbellVignjevic2000}, it can be observed that nonphysical penetration of the high explosive particles occurs. For the proposed contact method, the contact between the projectile and the sleeve, the projectile and the high explosives, and the high explosives and the sleeve are well simulated and no material penetration through the interface occurs.

\begin{figure}[htbp]
\centering
\includegraphics[width=16cm]{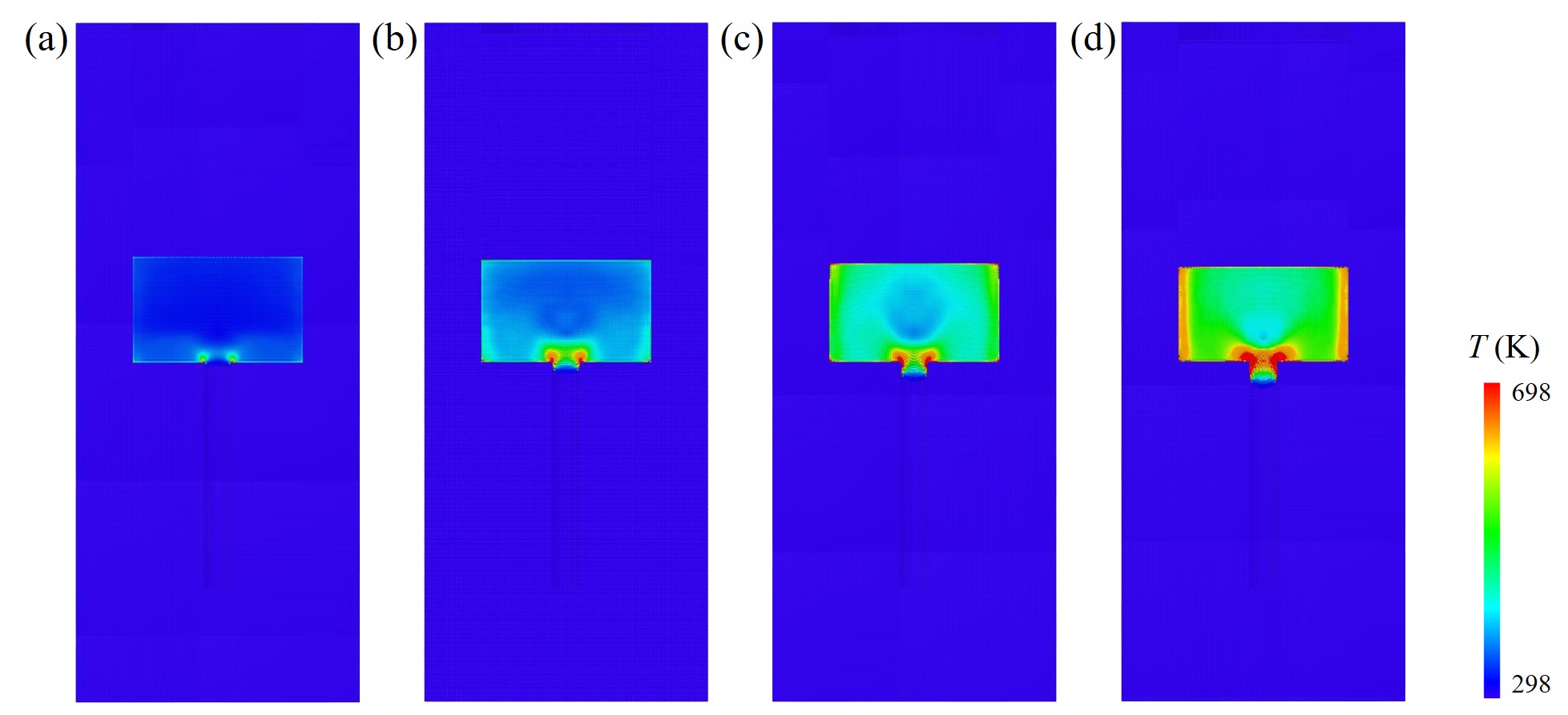}
\captionsetup{font={normalsize,stretch=1.0}}
\caption{Evolution of hot spots temperature associated with the microcracks under crack extrusion loading at (a) 8 $ \upmu \rm s $; (b) 12 $ \upmu \rm s $; (c) 16 $ \upmu \rm s $; (d) 20 $ \upmu \rm s $.}
\label{fig:Figure-4-2-4}
\end{figure}

\begin{figure}[htbp]
\centering
\includegraphics[width=10cm]{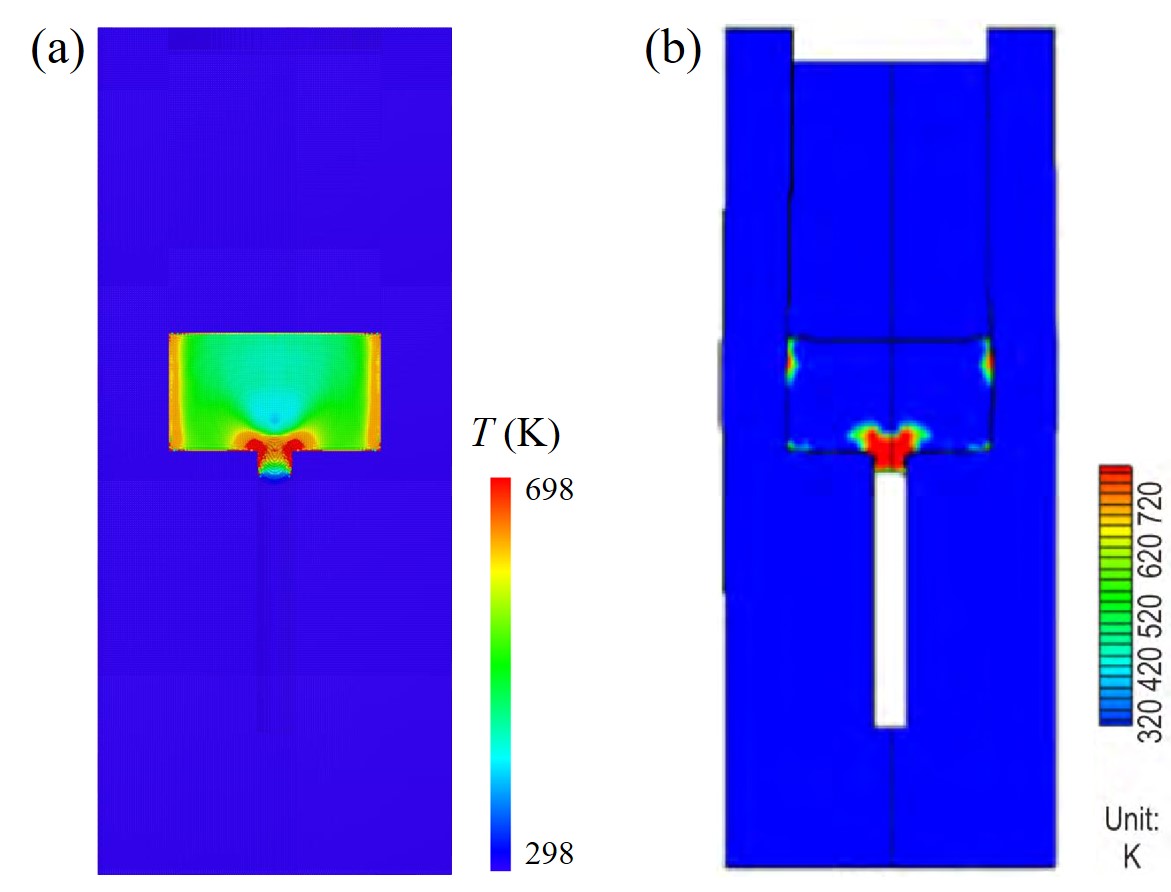}
\captionsetup{font={normalsize,stretch=1.0}}
\caption{Comparison of hot spots temperatures. (a) Present method; (b) Finite element method method \cite{YangWu2020}.}
\label{fig:Figure-4-2-5}
\end{figure}

Figure~\ref{fig:Figure-4-2-4} shows the evolution of hot spots associated with the microcracks. At 8 $ \upmu \rm s $, the high explosive around the crack is subjected to shear friction, and the crack surface rapidly frictions and generates heat, further inducing the material to ignite first. With the increase of time, the ignition region of the material propagates in the direction of $ 45^{\circ} $ to the lower surface. The microcracks in the self-locking area formed in the triangle and the enclosed region above the circular crack do not propagate in shear, and the corresponding hot spot temperature does not reach the ignition temperature. In addition, the high explosives are compressed by the side boundaries and a temperature increase also occurs. The variation of the hot spot temperature is in close agreement with the finite element method results \cite{YangWu2020}, as shown in Fig.~\ref{fig:Figure-4-2-5}.

\subsection{High velocity impact}

High velocity impact refers to the type of impact where the intensity of the shock wave generated by the impact is much larger than the yield strength of the material, which involves the kinetic response and dynamic fragmentation of the material under high pressure, and the multi-phase mixing and transformation of solid, fluid, gas, and plasma. The study of high velocity impact has considerable application prospects in spacecraft safety, high velocity kinetic energy destruction technology, armor design, and Earth-like planet meteorite impacts \cite{Jing1990, ZhangChen2021}.

\begin{figure}[htbp]
\centering
\includegraphics[width=4.5cm]{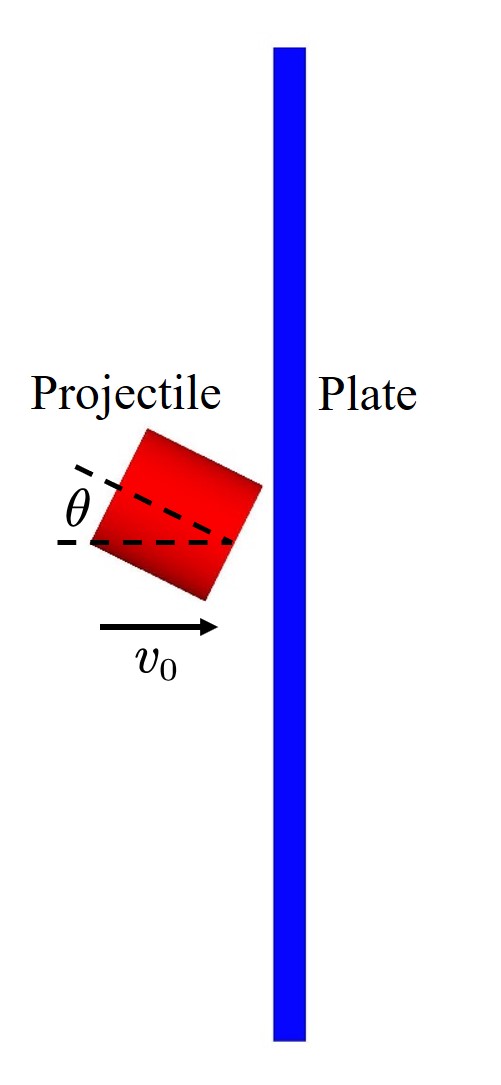}
\captionsetup{font={normalsize,stretch=1.0}}
\caption{Computational model of a cylindrical projectile impacting a target plate.}
\label{fig:Figure-4-3-1}
\end{figure}

Figure~\ref{fig:Figure-4-3-1} shows the computational model of a cylindrical projectile impacting a target plate. The length of the cylindrical projectile is 7.8 mm and its diameter is 7.8 mm. The length, width, and thickness of the plate are 60.0 mm, 60.0 mm, and 2.0 mm, respectively. The projectile had an angle of attack of $ 24.3^{\circ} $ and an initial velocity of 6.39 km/s in the $ x $ direction. The projectile material is Al 2024 and the target plate material is Al 6061. Both aluminum alloys are used with the Mie-Gr$\rm {\ddot{u}}$neisen equation of state and the Johnson-Cook constitutive model. The parameters of Al 2024 and Al 6061 are shown in Tables~\ref{tab:Table-4-3-1} and~\ref{tab:Table-4-3-2}. The initial particle distance is 0.1 mm and the total number of particles is 7573776. In the numerical simulations, the plate surface is selected as the master surface. Because of the large deformations that occur in the plate, it is necessary to reconstruct the local surface of the plate where contact occurs at each time step.

\begin{table}[htbp]
\centering
\caption{Johnson-Cook model parameters for Al 2024}
\begin{tabular}
{p{23mm}<{\centering}
 p{23mm}<{\centering}
 p{23mm}<{\centering}
 p{23mm}<{\centering}
 p{23mm}<{\centering}
 p{23mm}<{\centering}
}
\toprule

 $ \rho_0 \left( \rm kg / cm^3 \right) $ &
 $ E\left( \mathrm{MPa} \right) $ &
 $ \nu $ &
 $ A\left( \mathrm{MPa} \right) $ &
 $ B\left( \mathrm{MPa} \right) $ &
 $ n $ \\

\midrule

 2780.0 &
 73083.0 &
 0.33 &
 369.0 &
 684.0 &
 0.73 \\

\midrule

 $ c $ &
 $ m $ &
  &
  &
  &
 \\

\midrule

 0.0083 &
 1.7 &
  &
  &
  &
 \\

\bottomrule
\end{tabular}
\label{tab:Table-4-3-1}
\end{table}

\begin{table}[htbp]
\centering
\caption{Johnson-Cook model parameters for Al 6061}
\begin{tabular}
{p{23mm}<{\centering}
 p{23mm}<{\centering}
 p{23mm}<{\centering}
 p{23mm}<{\centering}
 p{23mm}<{\centering}
 p{23mm}<{\centering}
}
\toprule

 $ \rho_0 \left( \rm kg / cm^3 \right) $ &
 $ E\left( \mathrm{MPa} \right) $ &
 $ \nu $ &
 $ A\left( \mathrm{MPa} \right) $ &
 $ B\left( \mathrm{MPa} \right) $ &
 $ n $ \\

\midrule

 2704.0 &
 71000.0 &
 0.33 &
 324.1 &
 113.8 &
 0.42 \\

\midrule

 $ c $ &
 $ m $ &
  &
  &
  &
 \\

\midrule

 0.002 &
 1.34 &
  &
  &
  &
 \\

\bottomrule
\end{tabular}
\label{tab:Table-4-3-2}
\end{table}

\begin{figure}[htbp]
\centering
\includegraphics[width=16cm]{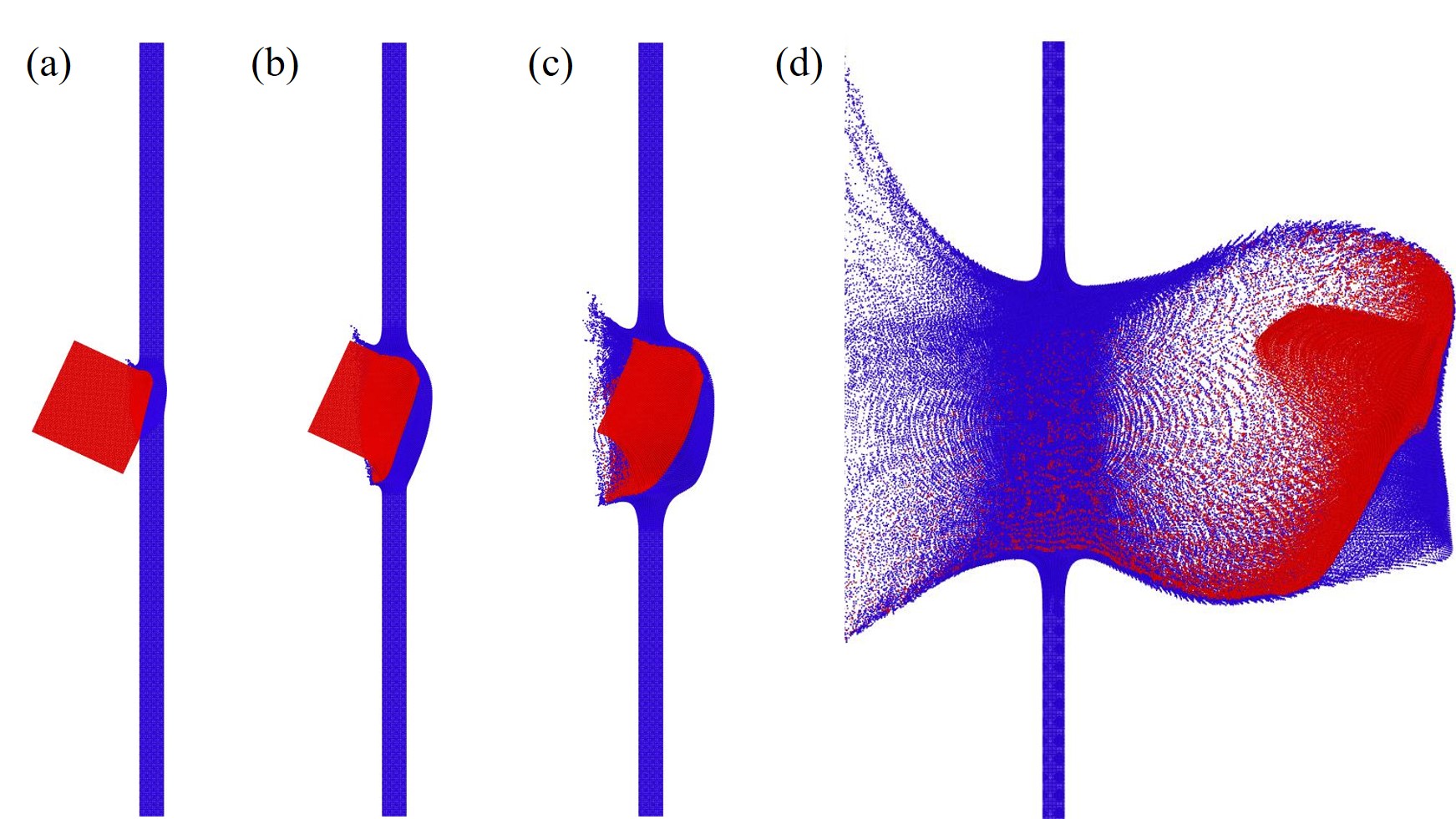}
\captionsetup{font={normalsize,stretch=1.0}}
\caption{The evolution of the debris cloud. (a) 0.4 $ \upmu \rm s $; (b) 0.8 $ \upmu \rm s $; (c) 1.2 $ \upmu \rm s $; (d) 6.0 $ \upmu \rm s $.}
\label{fig:Figure-4-3-2}
\end{figure}

\begin{figure}[htbp]
\centering
\includegraphics[width=14cm]{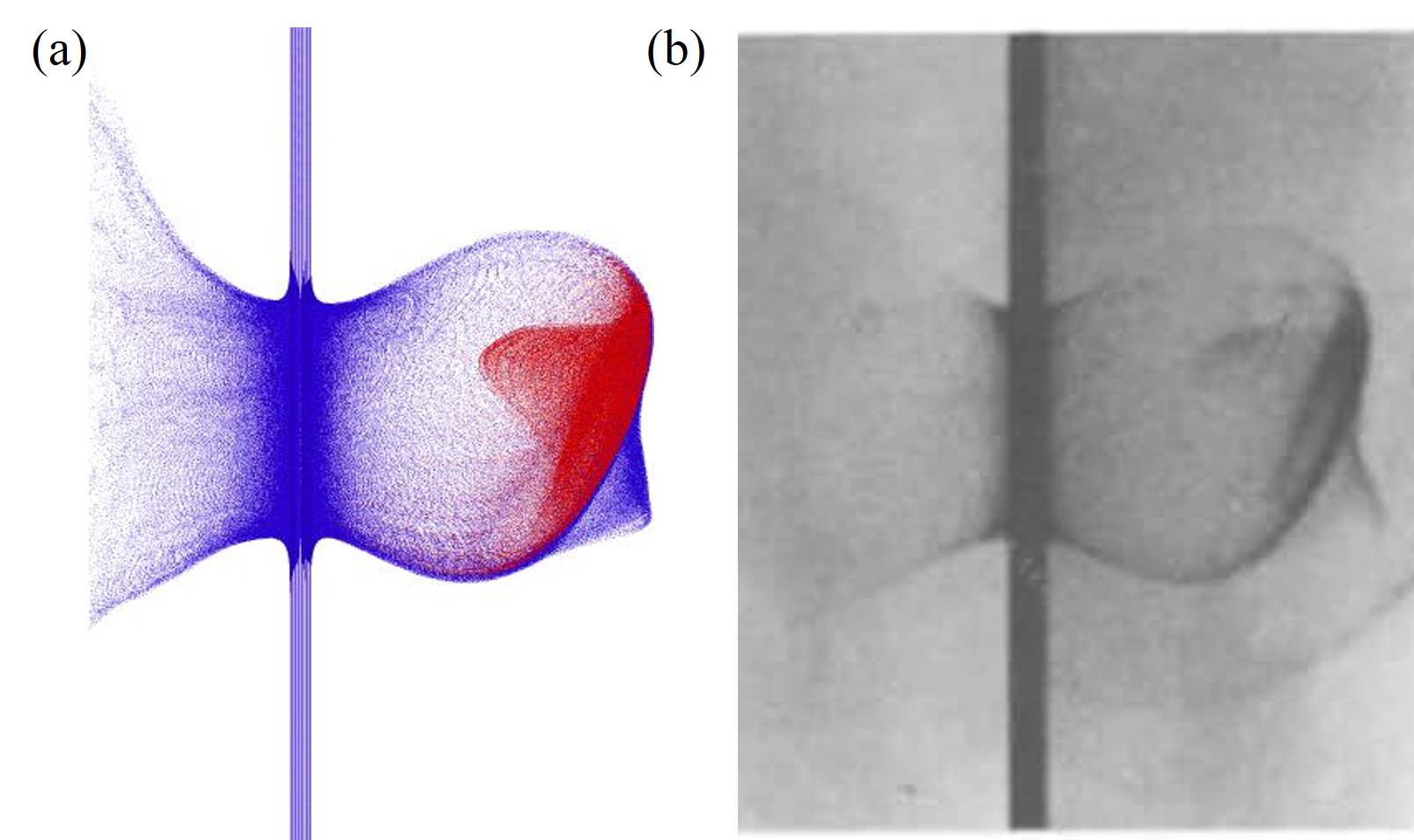}
\captionsetup{font={normalsize,stretch=1.0}}
\caption{Comparison of simulation result with experimental result \cite{Piekutowski1987}. For direct comparison with experiments, the simulation result is rotated $ 36.4^{\circ} $ clockwise along the $ x $-axis.}
\label{fig:Figure-4-3-3}
\end{figure}

Figure~\ref{fig:Figure-4-3-2} presents the evolution of the debris cloud, which includes the formation and expansion stages. In the formation stage, there is a strong interaction between the projectile and the plate, accompanied by intense energy transfer and conversion. In the expansion stage, the structure of the debris cloud is approximately unchanged, showing self-similar evolution, and forming a stable expansion velocity and structure. Due to the large attack angle of the projectile, the part of the projectile that first contacts the plate has a strong shearing effect with the plate and generates a large number of fragments. Numerical simulations and experimental results of the debris cloud at the late stage of the impact are further given in Fig.~\ref{fig:Figure-4-3-3}. The overall profile shows that the simulation results agrees with the experimental results. The maximum velocity of the debris cloud is 5.75 km/s. The deviation of this value from the experimental data of 5.60 km/s is 2.7\%. The numerical maximum velocity of the debris cloud agrees with the experimental value \cite{Piekutowski1987}.

\begin{figure}[htbp]
\centering
\includegraphics[width=10cm]{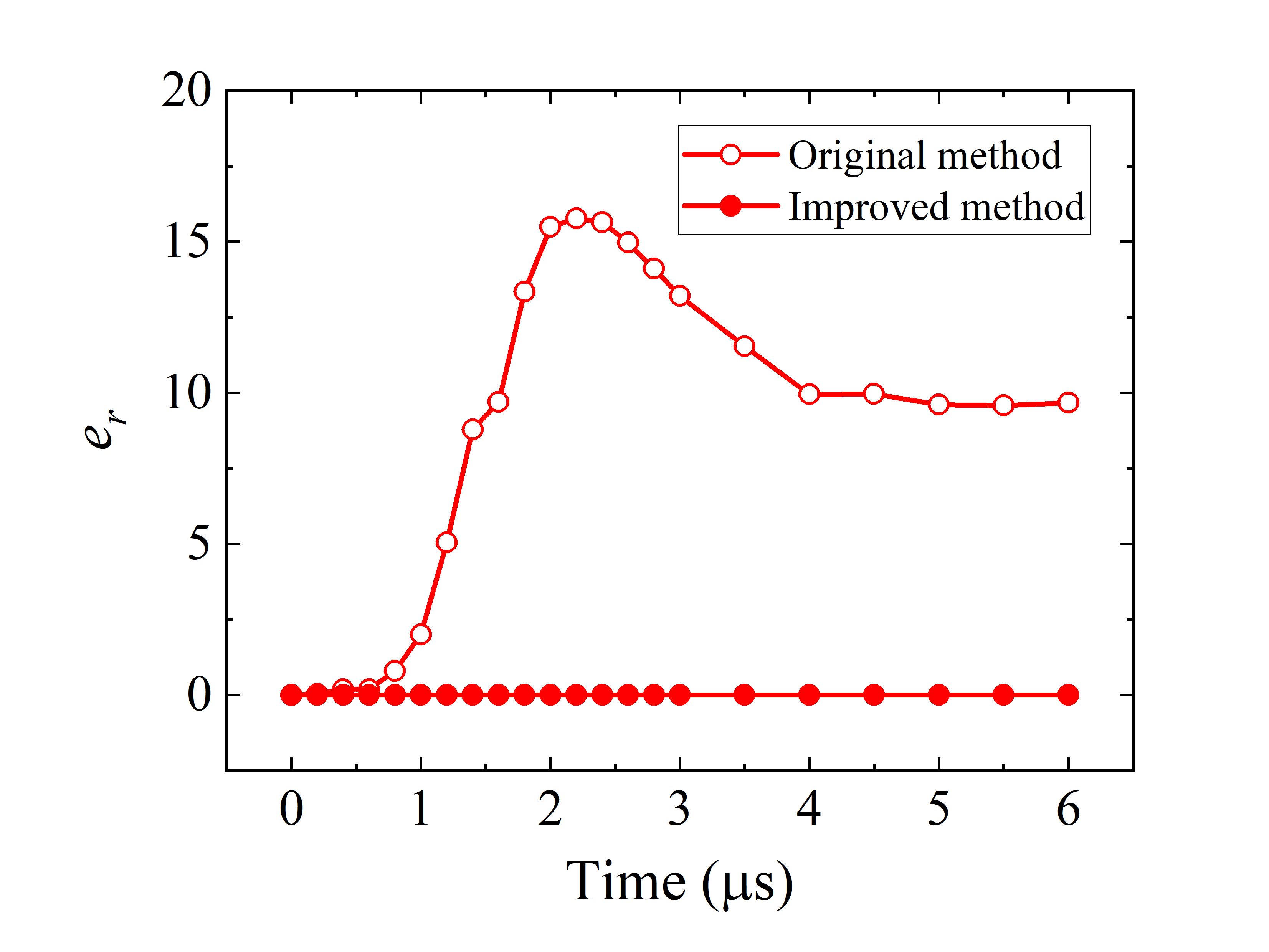}
\captionsetup{font={normalsize,stretch=1.0}}
\caption{The proportion of misidentified particles $ e_r = n_m / n_r $ using the original \cite{LiuMa2023} and improved methods over time, where $ n_r $ denotes the number of free-surface particles detected using the purely geometric method, and $ n_m $ denotes the number of misidentified particles.}
\label{fig:Figure-4-3-4}
\end{figure}

\begin{figure}[htbp]
\centering
\includegraphics[width=10cm]{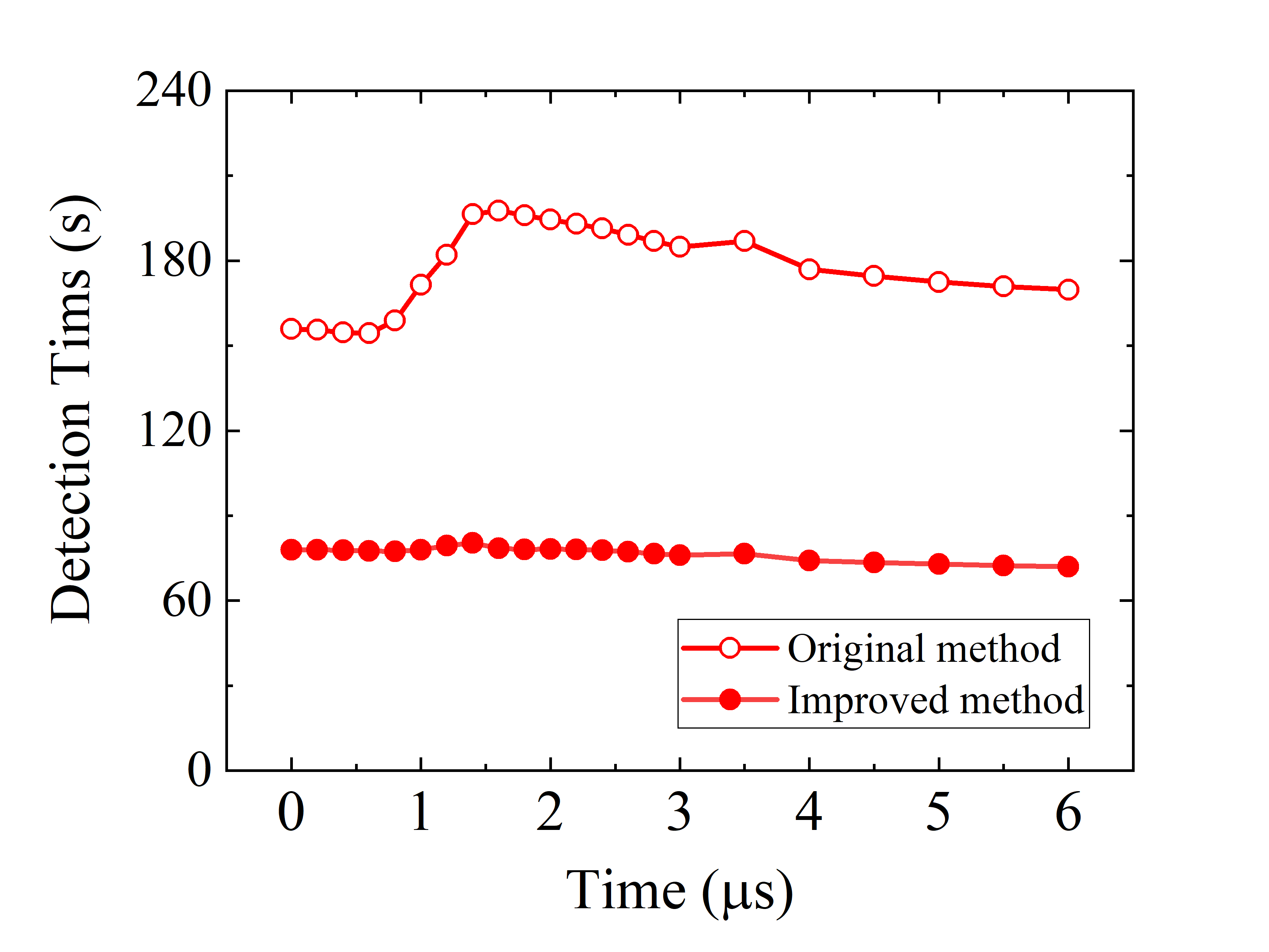}
\captionsetup{font={normalsize,stretch=1.0}}
\caption{The detection time over time using the original \cite{LiuMa2023} and improved methods.}
\label{fig:Figure-4-3-5}
\end{figure}

To analyze the improved free-surface particle detection method, free-surface particles at different moments are detected and compared with the original method \cite{LiuMa2023}. The detections are conducted on an i7-12700 processor with a main frequency of 2.10 GHz. Figure~\ref{fig:Figure-4-3-4} gives the results of the proportion of misidentified particles $ e_r = n_m / n_r $ using the original and improved methods over time, where $ n_r $ denotes the number of free-surface particles detected using the purely geometric method, and $ n_m $ denotes the number of misidentified particles. The results show that all free-surface particles can be accurately detected in the improved method. However, free-surface particles are misidentified up to 15.78\% in the original method because the effect of material compressibility is not considered. Figure~\ref{fig:Figure-4-3-5} gives the detection time over time using the original and improved methods. The results show that the improved method significantly reduces the detection time due to the addition of the detection along the normal vector in the second step.

\begin{figure}[htbp]
\centering
\includegraphics[width=16cm]{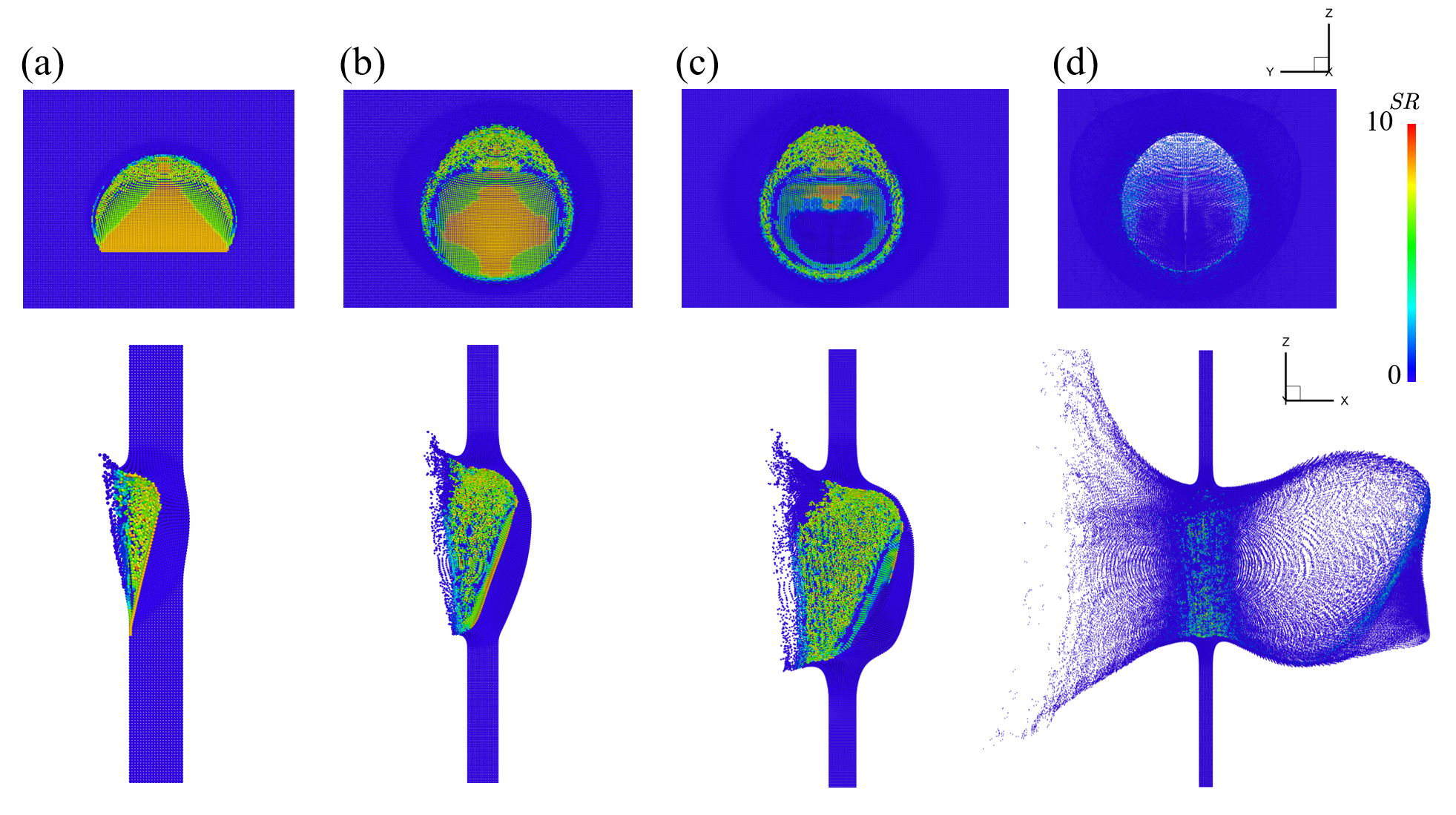}
\captionsetup{font={normalsize,stretch=1.0}}
\caption{Results of reconstructed local surface of free-surface particles on the plate surfaces. Since the local surface reconstruction of different free-surface particles is independent of each other and their surface triangles overlap, the overall reconstructed surface cannot be given, but the number of surface triangles $ SR $ for free-surface particles is given. (a) 0.4 $ \upmu \rm s $; (b) 0.8 $ \upmu \rm s $; (c) 1.2 $ \upmu \rm s $; (d) 6.0 $ \upmu \rm s $.}
\label{fig:Figure-4-3-6}
\end{figure}

\begin{figure}[htbp]
\centering
\includegraphics[width=10cm]{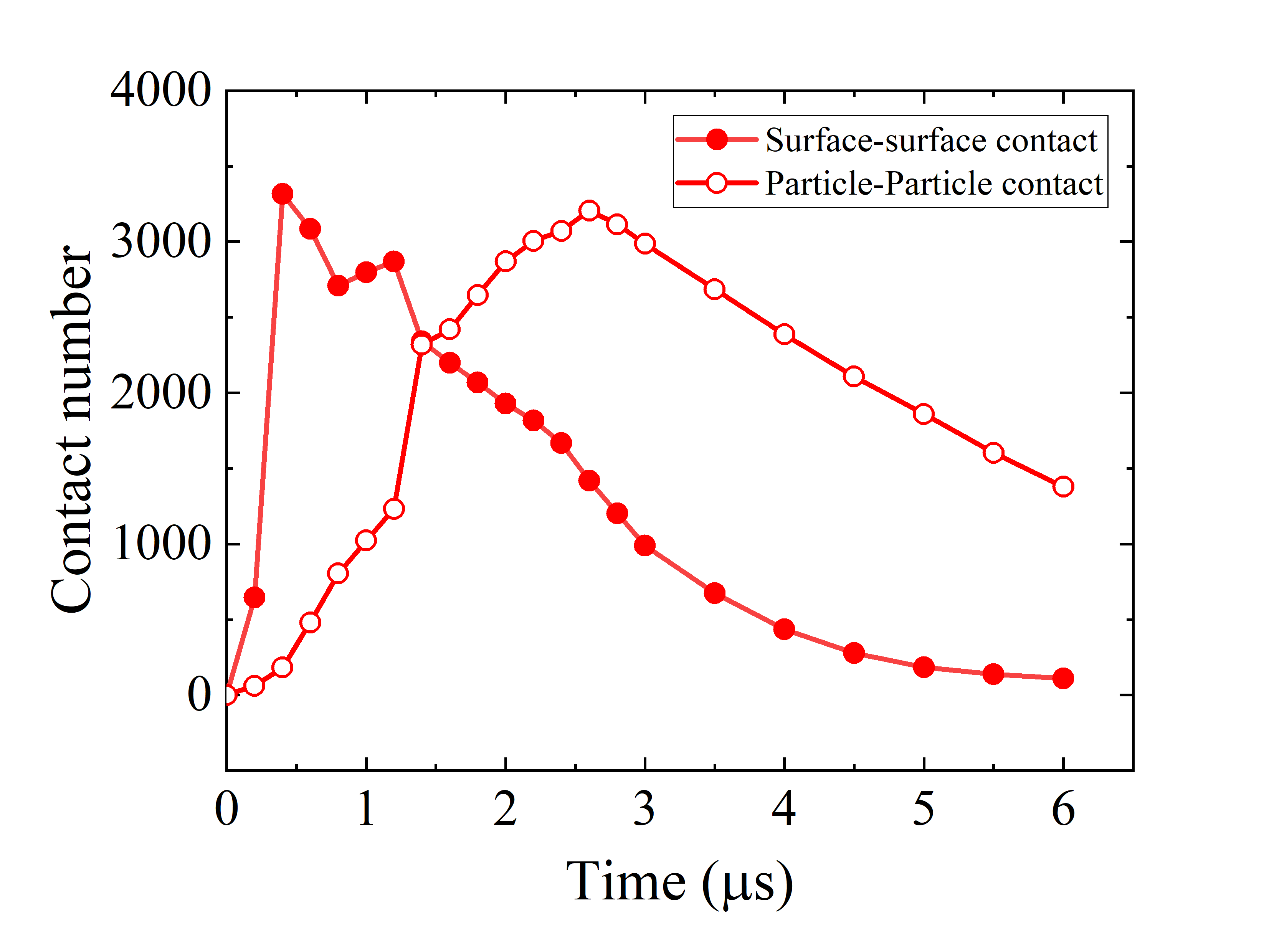}
\captionsetup{font={normalsize,stretch=1.0}}
\caption{The surface-surface contact and particle-particle contact numbers over time.}
\label{fig:Figure-4-3-7}
\end{figure}

Figures~\ref{fig:Figure-4-3-6} and~\ref{fig:Figure-4-3-7} give the results of local surface reconstruction of the plate surface (master surface) at different moments. When the projectile approaches the target plate, the particles on the plate that may be in contact with the particles of the projectile need to reconstruct the local surface, and the computation time is much less than the global surface reconstruction because only a few particles need to reconstruct the local surface. With the further evolution of the debris cloud, the deformation of the target plate becomes larger, and some free-surface particles cannot reconstruct the local surface, and the contact of these particles gradually develops from surface-surface contact to particle-particle contact. At 6.0 $ \upmu \rm s $, it can be seen that most of the particles in contact with the projectile are difficult to reconstruct the local surface, and at this time, the particle-particle contact will play a key role in the later evolution of the debris cloud. Therefore, it is necessary to supplement the particle-particle contact method with the surface-surface contact method in the extreme deformation numerical simulation.

\section{Conclusions}\label{sec:5}

In this study, we propose a generalized 3D hybrid contact method for SPH. First, an improved high accuracy free-surface particle detection method is developed. Then, a novel 3D local surface reconstruction method is developed based on the free-surface particles and region growing method, followed by the surface-surface contact detection and enforcement of normal penalty and tangential friction forces. Finally, the particle-particle contact method is added to deal with cases where surface-surface contact fails. The proposed method is first tested by simulating a block sliding along a slope and subsequently employed to simulate the ignition response of high explosives and high velocity impact. The conclusions are as follows.

(1) For the improved high accuracy free-surface particle detection method, the detection time is greatly reduced due to the addition of a single cone detection along the normal direction in the second step, and the detection method is successfully applied to the compressible field with good results due to the consideration of the effect of material density.

(2) The developed local surface reconstruction method has achieved good results for material surface reconstruction under various deformation situations, and it facilitates parallel processing because the local surface reconstruction of each particle is carried out independently.

(3) The results show that the proposed 3D hybrid contact method can achieve high accuracy and stability and is capable of handling various contact problems, including small, large, and extreme deformation problems.

\section*{Acknowledgments}

This work was supported by the China Postdoctoral Science Foundation (Grant No. 2023M740256).

\section*{Reference}

\bibliographystyle{elsarticle-num}
\bibliography{ConAlgorithm}

\end{document}